\documentclass[12pt]{article}

\usepackage{amsmath,amsthm,amssymb} 
\usepackage{mathbbol} 
\usepackage{enumitem} 
\usepackage{graphicx} 
\usepackage{xcolor} 

\usepackage{hyperref} 

\usepackage[labelformat=simple]{subcaption} 

\usepackage{authblk}

\theoremstyle{definition}

\title{Logarithmic lattice models for flows with boundaries}
\author[1]{Ciro S. Campolina}
\author[2]{Alexei A. Mailybaev}
\affil[1]{Universit\'{e} C\^{o}te d’Azur, CNRS, Inria, Institut de Physique de Nice, France\thanks{\texttt{ciro.sobrinho-campolina-martins@inria.fr}}}
\affil[2]{Instituto de Matem\'{a}tica Pura e Aplicada -- IMPA, Rio de Janeiro, Brazil}

\date{}

\begin{document}
	
	\maketitle
	
	\begin{abstract}
		Many fundamental problems in fluid dynamics are related to the effects of solid boundaries.
		In general, they install sharp gradients and contribute to the developement of small-scale structures, which are computationally expensive to resolve with numerical simulations.
		A way to access extremely fine scales with a reduced number of degrees of freedom is to consider the equations on logarithmic lattices in Fourier space.
		Here we introduce new toy models for flows with walls, by showing how to add boundaries to the logarithmic lattice framework.
		The resulting equations retain many important properties of the original systems, such as the conserved quantities, the symmetries and the boundary effects.
		We apply this technique to many flows, with emphasis on the inviscid limit of the Navier-Stokes equations.
		For this setup, simulations reach impressively large Reynolds numbers and disclose interesting insights about the original problem.
	\end{abstract}
	
	\section{Introduction}
	
	The presence of solid walls installs many subtleties in the dynamics of fluids.
	Boundaries naturally break homogeneity and act on the flow as sources of vorticity.
	Those aspects add extra complexity to fundamental mathematical and physical questions that are also formulated for boundaryless domains.
	
	An important problem, for instance, is whether Navier-Stokes smooth solutions converge to Euler's in the vanishing viscosity limit.
	While the convergence is well-established in full space~\cite{kato1972nonstationary}, the case with boundaries is still not well-understood~\cite{weinan2000boundary}.
	At high Reynolds numbers, viscous effects and sharp gradients are confined to a region called boundary layer~\cite{schlichting1961boundary}, which is supposed to remain in a vicinity of the wall, but might detach from it.
	The propagation of vortices generated on the boundary to the bulk flow has been linked to a possible anomalous dissipation in the inviscid limit.
	This would prevent the convergence of Navier-Stokes solutions to Euler's~\cite{kato1984remarks} and has been object of numerical investigations~\cite{nguyen2011energy,nguyen2018energy}.
	
\begin{figure}[hp]
		\begin{subfigure}[b]{0.475\textwidth}
			\centering
			\caption{}
			\includegraphics[width=\textwidth]{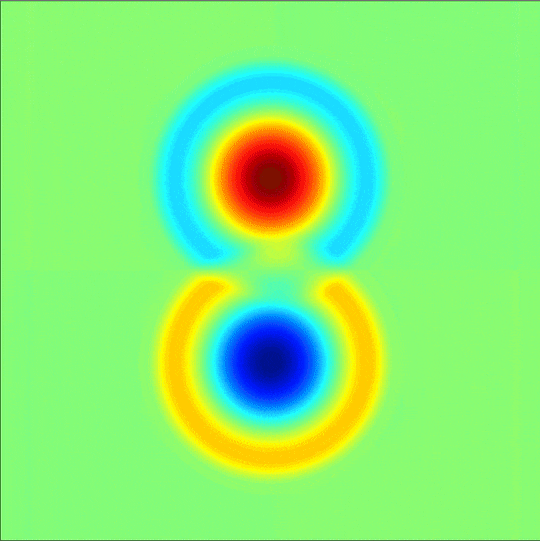}
		\end{subfigure}
		\hfill
		\begin{subfigure}[b]{0.475\textwidth}
			\centering
			\caption{}
			\includegraphics[width=\textwidth]{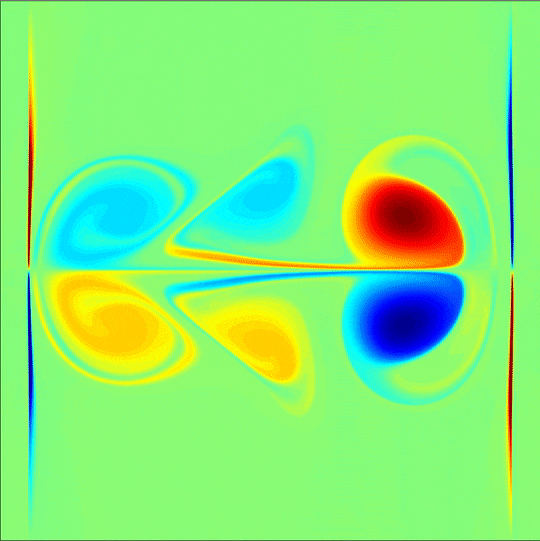}
		\end{subfigure}
		\vskip\baselineskip
		\begin{subfigure}[b]{0.475\textwidth}
			\centering
			\caption{}
			\includegraphics[width=\textwidth]{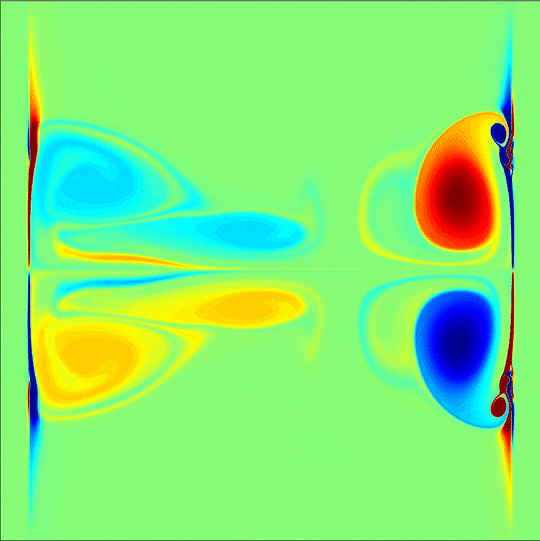}
		\end{subfigure}
		\hfill
		\begin{subfigure}[b]{0.475\textwidth}
			\centering
			\caption{}
			\includegraphics[width=\textwidth]{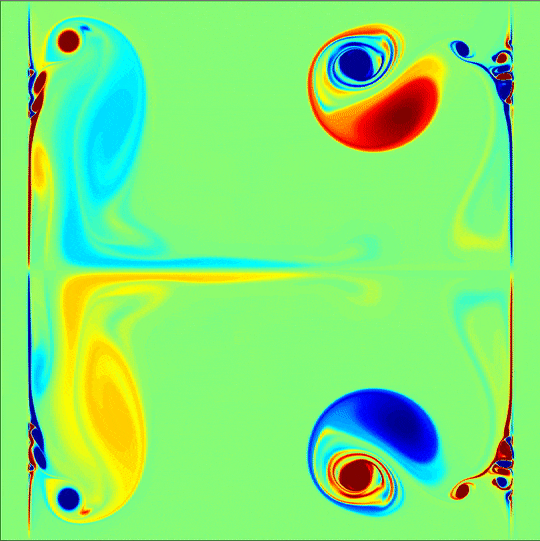}
		\end{subfigure}
		\hfill
		\caption{Direct numerical simulations of a dipole vortex colliding with a solid wall.
			Snapshots are for vorticity field.
			The flow is confined by two solid boundaries on left and right.
			Periodic boundary conditions are imposed on top and bottom.
			The Navier-Stokes equations are integrated using a pseudospectral method~\cite{trefethen2000spectral} and the solid boundary is modeled by a volume penalization method~\cite{nguyen2011energy}.
			We time-step with a classical 4th order Runge-Kutta scheme using an integration factor.
			Time evolves from (a) to (d).}
		\label{FIG:DNS_dipole}
	\end{figure}

	An archetype that has been considered in computational simulations is the case of dipole-wall collisions~\cite{kramer2007vorticity}.
	Their dynamics can be visualized in the vorticity snapshots in Fig.~\ref{FIG:DNS_dipole}.
	The flow is confined between two solid boundaries from left and right.
	We impose periodic boundary conditions in the top and the bottom.
	We initialize the flow with a dipole of opposite vorticities in the center of the domain---Fig.~\ref{FIG:DNS_dipole}(a).
	After the initialization, the cores of the dipole drift towards right.
	A sharp vorticity strip is developed very close to the bondaries---Fig.~\ref{FIG:DNS_dipole}(b).
	This is Prandtl boundary layer, which becomes closer and closer to the boundary as viscosity decreases~\cite{schlichting1961boundary}.
	When the dipole collides with the wall, small but intense vorticity structures arise from the boundary---Fig.~\ref{FIG:DNS_dipole}(c).
	They have the opposite sign of the vortex.
	This sharp adverse vorticity field arising from the walls is usually known as the boundary layer detachment.
	As a consequence, the large scale flow of the dipole rebound away from the wall---Fig.~\ref{FIG:DNS_dipole}(d)---, while small scale structures continue to be generated at the boundaries.
	All this phenomenology of boundary layers is strongly tight to the no-slip boundary condition combined with a very small viscosity.
	The inviscid case, however, is completely different.
	Euler's flow cannot penetrate boundaries, but it can slip along them, and this is what happens to inviscid solutions---consult~\cite[pp.~712-713]{nguyen2018energy} for snapshots of this same dipole-wall collision but evolved by Euler equations.
	These two contrasting states after collision---rebounded vortices for Navier-Stokes and its slippery Euler counterpart---strongly indicate a possible lack of convergence of the inviscid limit.
	
	Despite the relevant numerical investigations, the mechanism of unsteady detachment is still under debate~\cite{cassel2000comparison,weinan2000boundary}, while the Prandtl boundary layer equations also manifest a rich variety of features, such as instabilities~\cite{grenier2000nonlinear}, finite-time singularities~\cite{weinan1997blowup,kukavica2017van}, and ill-posedness~\cite{gerard2010ill}.
	For a review of these and other interesting mathematical problems surrounding the effects of boundaries we refer the reader to~\cite{bardos2013mathematics,lopes2018fluids}.
	
	To overcome computational limitations, toy models are traditionally employed in order to investigate physical phenomena.
	Good example are the \textit{shell models}~\cite{biferale2003shell}, which have been successfully used to study important aspects of turbulence of homogeneous (boundaryless) flows.
	The subtle question lies on which features of the original model should be kept and which should be dropped.
	A framework recently proposed for the formulation of toy models is the \textit{logarithmic lattice}~\cite{campolina2021fluid}.
	In this configuration, we consider a reduced number of degrees of freedom by taking the governing equations on logarithmic lattices in Fourier space.
	This allows to access etremely fine scales of the flow.
	Such technique was successfully applied to the study of the chaotic blowup in the 3D incompressible Euler equations~\cite{campolina2018chaotic} and to Navier-Stokes turbulence~\cite{campolina2021fluid}, among other applications~\cite{barral2023asymptotic,costa2023reversible,pikeroen2023log,pikeroen2023tracking}.
	Nevertheless, all the studies so far considered only homogeneous flows, that is, flows without boundaries.
	
	In this paper, we present some toy models for flows with boundaries, by showing how to introduce boundaries on the logarithmic lattice framework.
	The key challenge resides on how to model walls and restricted domains with Fourier variables.
	We do it in a systematic way by extending the flow to the whole Euclidean space through symmetrization.
	This process introduces some jump singularities accross the boundaries, that need to be carefully treated.
	The resulting model retains the same properties of the original equations, such as the conserved quantities, the symmetry groups and the boundary effects.
	With this model, we investigate the inviscid limit problem of the Navier-Stokes equations.
	Reynolds number can be raised to impressivevily large values and we provide some insights on this long-standing problem.
	
	The paper is organized as follows.
	In Section~\ref{SEC:solid_singular} we explain how to add boundaries on logarithmic lattices.
	The full 3D incompressible Navier-Stokes equations with no-slip walls are systematically derived on this framework.
	In Section~\ref{SEC:1Dshear} we consider classical shear flows, which provide great intuition and rigor on the unusual understanding of boundaries on Fourier variables.
	In Section~\ref{SEC:2DBL} we address the inviscid limit problem of the Navier-Stokes equations in two dimensions.
	In Section~\ref{SEC:CONC} we draw some conclusions.
	
	
	\section{Boundaries on logarithmic lattices}\label{SEC:solid_singular}
	
	Logarithmic lattice fields are inherently defined on Fourier space, which is not readily available for physical domains other than the full Euclidean space.
	For this reason, it is chalenging to consider solid boundaries and restricted domains on logarithmic lattices.
	Our strategy is to extend the flow to the whole space and then to model the boundaries as surfaces immersed in the fluid.
	In this process, discontinuities might incur accross the boundary, and so we need to consider a discontinuous formulation of the governing equations~~\cite{Sirovich1967Ini,Goldstein1993Mod}.
	This approach is inspired by the well-known immersed boundary method~\cite{peskin1972flow,peskin2002immersed}.
	
	
	\subsection{Fluid dynamics equations with immersed boundary}
	
	We consider a three-dimensional velocity field $\mathbf{u}(x,y,z) = (u,v,w)$ in the upper-half space $y>0$ with a steady solid boundary on the plane $y = 0$.
	Viscous incompressible flow in this domain is governed by the classical Navier-Stokes equations together with no-slip boundary condition
	\begin{equation}\label{EQ:3D_flow}
		\begin{cases}
			\partial_t \mathbf{u} + \mathbf{u} \cdot \nabla \mathbf{u} = -\nabla p + \nu \Delta \mathbf{u} \quad &\text{in} \ y >0,\\
			\nabla \cdot \mathbf{u} = 0 \quad &\text{in} \ y >0,\\
			\mathbf{u} = \mathbf{0} \quad &\text{on} \ y = 0.
		\end{cases}
	\end{equation}
	System~\eqref{EQ:3D_flow} also approximates the governing set of equations for smooth boundaries of more general geometries, when considering the flow in a small vicinity of a boundary point taken as the origin in local Cartesian coordinates.
	
\begin{figure*}[t]
		\centering
		\includegraphics[width=.85\textwidth]{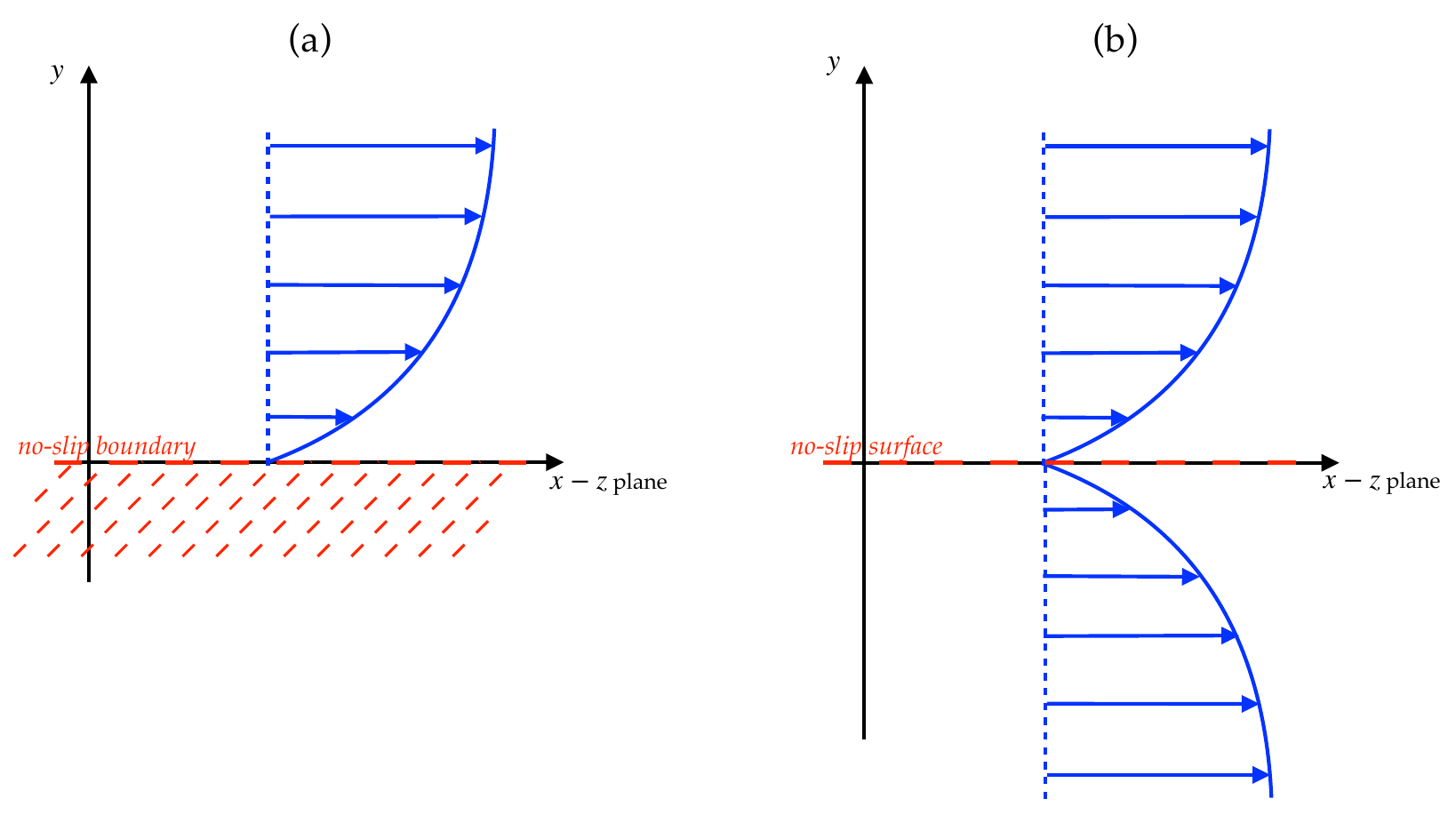}
		\caption{Extension of (a)~a flow restricted to the upper-half space $y>0$ with a no-slip boundary at $y=0$ to (b) a flow defined on the whole Euclidean space $\mathbb{R}^3$ through symmetries~\eqref{EQ:flow_symmetry} with a no-slip immersed surface at its original place $y=0$.}
		\label{FIG:reflection_at_boundary}
	\end{figure*}

	Our goal is to deduce a system equivalent to~\eqref{EQ:3D_flow}, but with the new field variables $u$, $v$, $w$ and $p$ defined everywhere.
	A simple way to achieve this is by extending the flow to the lower-half space $y<0$ through the symmetries
	\begin{equation}
		\begin{alignedat}{3}
			u(x,y,z,t) &= &u(x,-y,z,t), \\
			v(x,y,z,t) &= -&v(x,-y,z,t), \\
			w(x,y,z,t) &= &w(x,-y,z,t), \\
			p(x,y,z,t) &= &p(x,-y,z,t),
		\end{alignedat}
		\label{EQ:flow_symmetry}
	\end{equation}
	for all $x$, $y$, $z$ and $t$.
	Fig.~\ref{FIG:reflection_at_boundary} illustrates this symmetrization.
	Observe that, because of the parity of $v$, any symmetric field~\eqref{EQ:flow_symmetry} immediately satisfies the no-penetration condition on the boundary
	\begin{equation}
		v = 0 \ \text{at} \ y=0.
	\end{equation}
	However, in performing such extension, we might have introduced singularities in the system.
	More precisely, the resulting reflected field variables may present jump discontinuities in their derivatives accross the solid boundary.
	Therefore, we should consider the axis $y = 0$ as a steady discontinuity surface, and the balance laws must take the jump singularities into account.
	The governing equations we should consider are the \textit{discontinuous formulation of the Navier-Stokes equations}~\cite{Sirovich1967Ini}
	\begin{equation}\label{EQ:flat_plate_model}
		\begin{cases}
			\partial_t \mathbf{u} + \mathbf{u} \cdot \nabla \mathbf{u} = -\nabla p + \nu \Delta \mathbf{u} -\nu \mathbf{J}(x,z,t)\delta(y)\quad &\text{in} \ \mathbb{R}^3,\\
			\nabla \cdot \mathbf{u} = 0 \quad &\text{in} \ \mathbb{R}^3,\\
			\mathbf{u} = \mathbf{0} \quad &\text{on} \ y = 0,\\
			\text{+ symmetries~\eqref{EQ:flow_symmetry}}.
		\end{cases}
	\end{equation}
	where $\mathbf{J}(x,z,t)$ is the jump of the shear velocities accross the boundary
	\begin{equation}\label{EQ:J}
		\mathbf{J} =
		\begin{pmatrix}
			[\partial_y u]\\
			0\\
			[\partial_y w]
		\end{pmatrix}.
	\end{equation}
	Here, the square brackets $[\ \cdot \ ]$ represent the jump of a field accross the boundary $[f] = f(y=0^+) - f(y=0^-)$, and $\delta(y)$ is the Dirac delta distribution.
	Note that, due to the parity~\eqref{EQ:flow_symmetry} of $v$, we have zero $y$-shear component $[\partial_y v] = 0$.
	We present details of the deduction of this model in Appendix~\hyperref[app:A]{A}.
	
	The concentrated force $-\nu \mathbf{J}(x,z,t)\delta(y)$ is the shear exerted by the solid plate on the fluid and is responsible for relaxing the flow velocity $\mathbf{u}$ (more precisely, components $u$ and $w$) to zero on the disontinuity surface $y = 0$.
	
	Just as pressure is obtained from incompressibility, the jump variable $\mathbf{J}$ is determined from the no-slip condition, as we are going to see for the logarithmic lattice formulation in the next subsection. A way to compute it in the continuous framework is suggested in~\cite{campolina2022fluid}.
	
	As we can see from the direct deduction of the models, the original boundary value problem~\eqref{EQ:3D_flow} and its immersed boundary formulation~\eqref{EQ:flat_plate_model} are equivalent.
	A more precise argument for such statement can be found in~\cite{campolina2022fluid}.
	
	
	
	We remark the necessity of imposing the no-slip condition on the discontinuity surface for the uniqueness of solutions.
	Indeed, the no-slip condition cannot be lifted up from the set of equations~\eqref{EQ:flat_plate_model}, otherwise the resulting system would be ill-posed---see Appendix~\hyperref[app:B]{B} for an example of nonunique solutions from the same initial data when no-slip condition is not explicitly prescribed.
	
	In the inviscid case $\nu = 0$, we simplify~\eqref{EQ:flat_plate_model} to the \textit{incompressible Euler equations}
	\begin{equation}\label{EQ:flat_plate_model_inviscid}
		\begin{cases}
			\partial_t \mathbf{u} + \mathbf{u} \cdot \nabla \mathbf{u} = -\nabla p \quad &\text{in} \ \mathbb{R}^3,\\
			\nabla \cdot \mathbf{u} = 0 \quad &\text{in} \ \mathbb{R}^3,\\
			\text{+ symmetries~\eqref{EQ:flow_symmetry}}.
		\end{cases}
	\end{equation}
	In this system, no penetration
	\begin{equation}
		v = 0 \quad \text{in} \ y = 0
	\end{equation}
	is satisfied as a consequence of the flow symmetry, while the flow may slip through the boundary, \textit{i.e.} $u \neq 0$ and $w \neq 0$ on $y = 0$.
	
	\subsection{Logarithmic lattice formulation}
	
	In the immmersed boundary formulation, the flow is defined in the whole Euclidean space.
	This allows us to readily work with Fourier variables, and so on logarithmic lattices.
	
	We shall consider a three-dimensional logarithmic lattice $\mathbb{\Lambda}^3$, where
	\begin{equation}
		\mathbb{\Lambda} = \{ \pm 1, \pm \lambda, \pm \lambda^2, \dots \},
	\end{equation}
	for some $\lambda>1$.
	The usual choices are $\lambda =2$, $\lambda = (1+\sqrt{5})/2 \approx 1.618$ (the golden ratio), and $\lambda = (\sqrt[3]{9 + \sqrt{69}} + \sqrt[3]{9-\sqrt{69}})/\sqrt[3]{18} \approx 1.325$ (the plastic number), which result in nontrivial nonlinear interactions among modes~\cite{campolina2021fluid}.
	This lattice mimics Fourier space with largest integral scale $L \sim 2\pi$ corresponding to $|\mathbf{k}| \sim 1$.
	
 	We represent the velocity field $\mathbf{u}(\mathbf{k},t) = (u,v,w) \in \mathbb{C}^3$ as a function of the wave vector $\mathbf{k} = (k_x,k_y,k_z) \in \mathbb{\Lambda}^3$ on the logarithmic lattice and the time variable $t\in \mathbb{R}$.
	Similarly, we have the scalar pressure $p(\mathbf{k},t) \in \mathbb{C}$.
	These and all field variables are supposed to satisfy the reality condition $\mathbf{u}(-\mathbf{k},t) = \overline{\mathbf{u}(\mathbf{k},t)}$, where the overbar indicates complex conjugation.
	
	In physical space, the flow is reflected with respect to the plane $y = 0$ as in~\eqref{EQ:flow_symmetry}.
	Accordingly, we demand the lattice fields to satisfy the corresponding symmetries
	\begin{equation}
		\begin{alignedat}{3}
			u(k_x,k_y,k_z,t) &= &u(k_x,-k_y,k_z,t), \\
			v(k_x,k_y,k_z,t) &= -&v(k_x,-k_y,k_z,t), \\
			w(k_x,k_y,k_z,t) &= &w(k_x,-k_y,k_z,t), \\
			p(k_x,k_y,k_z,t) &= &p(k_x,-k_y,k_z,t),
		\end{alignedat}
		\label{EQ:flow_symmetry_lattice}
	\end{equation}
	for all $k_x$, $k_y$, $k_z$ and $t$.
	
	We have also to consider the action of the shear force at the boundary $\mathbf{F} = (F_x,F_y,F_z)$ on the flow, which is related to the jump discontinuities $\mathbf{J} = (J_x,J_y,J_z)$ in the form
	\begin{equation}\label{EQ:jump_lattice}
		\mathbf{F}(\mathbf{k},t) = -\nu \mathbf{J}(k_x,k_z,t) \delta(k_y).
	\end{equation}
	Since the jump discontinuities in physical space occur at the plane $y = 0$, the corresponding lattice variable is a function independent of $k_y$, and thus depends on $k_x$, $k_z$ and $t$ only.
	In analogy with the Fourier transform of Dirac delta distribution, we take the lattice Dirac delta function $\delta(k_y)$ as unity
	\begin{equation}\label{EQ:Dirac_delta_lattice}
		\delta(k_y) = 1 \quad \text{for all} \ k_y \in \mathbb{\Lambda}.
	\end{equation}
	Such natural definition preserves some important properties of Dirac delta, like scaling invariance
	\begin{equation}
		\delta(\lambda k_y) = \delta(k_y)
	\end{equation}
	and parity
	\begin{equation}
		\delta(-k_y) = \delta(k_y).
	\end{equation}
	
	Moreover, Dirac delta~\eqref{EQ:Dirac_delta_lattice} on the lattice keeps similarity with a classical property of the original distribution:
	if $f(k_y)$ is a lattice function representing a function $F(y)$ in physical space, then its $k_y$-inner product against delta mimics the localization of $F$ on $y = 0$, since
	\begin{equation}
		(f,\delta)_{k_y} = \sum_{k_y \in  \mathbb{\Lambda}}f(k_y)\overline{\delta(k_y)}
		= \sum_{k_y \in  \mathbb{\Lambda}}f(k_y)
		\simeq \int \hat{F}(k_y) dk_y = F(y) \Big|_{y = 0}.
	\end{equation}
	Here the left-hand side of $\simeq$ corresponds to a logarithmic lattice representation $k_y \in \mathbb{\Lambda}$, and the right-hand side corresponds to a usual representation in continuous space $y \in \mathbb{R}$.
	This interpretation allows us to impose the no-slip boundary condition in the form
	\begin{equation}\label{EQ:no_slip_lattice}
		(\mathbf{u},\delta)_{k_y} = \mathbf{0} \quad \text{for all} \ k_x,k_z \in \mathbb{\Lambda}, t \in \mathbb{R},
	\end{equation}
	since we interpret
	\begin{equation}
		(\mathbf{u},\delta)_{k_y} \simeq \mathbf{u}\Big|_{y = 0}.
	\end{equation}
	Observe that the left expression in Eq.~\eqref{EQ:no_slip_lattice} is a function of $k_x$, $k_z$ and $t$.
	
	With all the above definitions, we can now establish the \textit{incompressible Navier-Stokes equations with a solid boundary on a logarithmic lattice}.
	We simply write the immersed boundary formulation of the Navier-Stokes equations~\eqref{EQ:flat_plate_model}, but considering the fields and operations on logarithmic lattices
	\begin{equation}\label{EQ:flat_plate_model_lattice}
		\begin{cases}
			\partial_t \mathbf{u} + \mathbf{u} \ast \nabla \mathbf{u} = -\nabla p + \nu \Delta \mathbf{u} -\nu \mathbf{J}(k_x,k_z,t) \delta(k_y)\quad &\text{in} \ \mathbb{\Lambda}^3,\\
			\nabla \cdot \mathbf{u} = 0 \quad &\text{in} \ \mathbb{\Lambda}^3,\\
			(\mathbf{u},\delta)_{k_y} = \mathbf{0} \quad &\text{for all} \ k_x,k_z,t,\\
			\text{+ symmetries~\eqref{EQ:flow_symmetry_lattice}}.
		\end{cases}
	\end{equation}
	The corresponding \textit{incompressible Euler equations with solid boundary on a logarithmic lattice} reads
	\begin{equation}\label{EQ:Euler_plate_lattice}
		\begin{cases}
			\partial_t \mathbf{u} + \mathbf{u} \ast \nabla \mathbf{u} = -\nabla p\quad &\text{in} \ \mathbb{\Lambda}^3,\\
			\nabla \cdot \mathbf{u} = 0 \quad &\text{in} \ \mathbb{\Lambda}^3,\\
			\text{+ symmetries~\eqref{EQ:flow_symmetry_lattice}}.
		\end{cases}
	\end{equation}
	In this case, the no-penetration boundary condition $(v,\delta)_{k_y} = 0$ is a consequence of the symmetry $v(-k_y) = -v(-k_y)$ on $v$ from~\eqref{EQ:flow_symmetry_lattice}.
	We recall that the derivatives on logarithmic lattices are given by the Fourier factors
	\begin{equation}
		\partial_j f(\mathbf{k}) = ik_j f(\mathbf{k}),
	\end{equation}
	and that the star $\ast$ product is the convolution on the logarithmic lattice
	\begin{equation}
		(f\ast g)(\mathbf{k}) = \sum_{\mathbf{p},\mathbf{q} \in \mathbb{\Lambda}^3, \\ \mathbf{p}+\mathbf{q}=\mathbf{k}}f(\mathbf{p})g(\mathbf{q}).
	\end{equation}
	We refer the reader to~\cite{campolina2021fluid} for a comprehensive description of the logarithmic lattice framework.
	
	Taking the divergence of the momentum equation in~\eqref{EQ:flat_plate_model_lattice} and invoking incompressibility, we obtain a Poisson equation for the pressure
	\begin{equation}\label{EQ:poisson_pressure_boundary}
		\Delta p = -\nabla \cdot (\mathbf{u} \ast \nabla \mathbf{u}) - \nu \nabla \cdot (\mathbf{J}(k_x,k_z,t) \delta(k_y)).
	\end{equation}
	This allows us to eliminate the pressure from system~\eqref{EQ:flat_plate_model_lattice}.
	Observe, however, that the pressure is written as a function of not only the velocities, but of the jumps as well.
	The additional term $- \nu \nabla \cdot (\mathbf{J}(k_x,k_z,t) \delta(k_y))$ in pressure's equation stands for the contribution of the boundary.
	
	\subsection{Computation of the jump lattice variable}
	
	Just as pressure is obtained from the incompressibility constraint, the jump variables are computed from the no-slip condition.
	We show here how such calculation can be evaluated.
	The idea relies on first truncating the lattice to later pass the limit on the truncation.
	The truncated system is a finite dimensional system, whose jump contribution can be readily computed.
	
	Let us truncate the lattice $\mathbb{\Lambda}$ by consering only frequencies smaller than the wave number $k_N = \lambda^{N-1}$ as
	\begin{equation}\label{EQ:truncated_lattice}
		\mathbb{\Lambda}_N = \{ \pm 1, \pm \lambda, \pm \lambda^2, \dots, \pm \lambda^{N-1} \}.
	\end{equation}
	Then, the Dirac delta on the lattice~\eqref{EQ:Dirac_delta_lattice} is automatically ``regularized'', since truncation turns it into a summable function.
	
	Then we consider the approximated system
	\begin{equation}\label{EQ:flat_plate_model_lattice_reg}
		\begin{cases}
			\partial_t \mathbf{u}^N + \mathbf{u}^N \ast \nabla \mathbf{u}^N = -\nabla p^N + \nu \Delta \mathbf{u}^N -\nu \mathbf{J}^N(k_x,k_z,t) \delta^N(k_y)\quad &\text{in} \ \mathbb{\Lambda}^3_N,\\
			\nabla \cdot \mathbf{u}^N = 0 \quad &\text{in} \ \mathbb{\Lambda}^3_N,\\
			(\mathbf{u}^N,\delta^N)_{k_y} = \mathbf{0} \quad &\text{for all} \ k_x,k_z,t,\\
			\text{+ symmetries~\eqref{EQ:flow_symmetry_lattice}}.
		\end{cases}
	\end{equation}
	We wrote the superscript $N$ to indicate all truncated variables, \textit{i.e.} those defined on the truncated lattice~\eqref{EQ:truncated_lattice}.
	
	To compute the truncated jumps $\mathbf{J}^N$, we take the $k_y$-inner product of the momentum equation with $\delta^N$.
	As a consequence of no-slip condition, the contribution of time variation vanishes, and we are lead to
	\begin{equation}\label{EQ:product_with_delta}
		(\mathbf{u}^N \ast \nabla \mathbf{u}^N, \delta^N)_{k_y} = (-\nabla p^N + \nu \Delta \mathbf{u}^N, \delta^N)_{k_y} + (-\nu \mathbf{J}^N(k_x,k_z,t) \delta^N(k_y),\delta^N)_{k_y}.
	\end{equation}
	Using the fact that the jumps $\mathbf{J}^N$ do not depend on $k_y$, one may write
	\begin{equation}\label{EQ:contribution_F}
		(-\nu \mathbf{J}^N(k_x,k_z,t) \delta^N(k_y),\delta^N)_{k_y} = -\nu \mathbf{J}^N(k_x,k_z,t) (\delta^N,\delta^N)_{k_y}.
	\end{equation}
	Because the approximated truncated $\delta^N$ is summable, the product $(\delta^N,\delta^N)_{k_y}$ appearing in Eq.~\eqref{EQ:contribution_F} is a well-defined positive number.
	Substitution of expression~\eqref{EQ:contribution_F} into Eq.~\eqref{EQ:product_with_delta} yields, after some manipulations,
	\begin{equation}
		\mathbf{J}^N(k_x,k_z,t) = \frac{1}{\nu (\delta^N,\delta^N)_{k_y}}(-\mathbf{u}^N \ast \nabla \mathbf{u}^N - \nabla p^N + \nu \Delta \mathbf{u}^N, \delta^N)_{k_y}.
	\end{equation}
	Because of the flow symmetries~\eqref{EQ:flow_symmetry_lattice}, we have $J_y^N = 0$, as expected.
	We remark, however, that this is not a closed formula for the regularized jumps $\mathbf{J}^N$, but an implicit equation.
	Indeed, the pressure must be solved from the Poisson equation~\eqref{EQ:poisson_pressure_boundary} in terms of both the velocities $\mathbf{u}^N$ and the jumps $\mathbf{J}^N$.
	The resulting equations can be solved explicitly for $\mathbf{J}^N$, but we omit here the laborious computations.
	In Sections~\ref{SEC:1Dshear} and~\ref{SEC:2DBL} we present the resulting explict formula for the simpler cases of one and two dimensions, respectively.
	
	To complete the computations, we just need to take the limit $N \to \infty$.
	
	\subsection{Basic symmetries and balance laws}

	The Euler equations on logarithmic lattices are known for having the same symmetries as the original equations~\cite{campolina2019fluid}.	
	When adding the boundary, the new restricted symmetry group should preserve the imposed flow symmetries~\eqref{EQ:flow_symmetry_lattice}.
	To enumerate them, if $\mathbf{u}(\mathbf{k},t)$ is a solution of system~\eqref{EQ:Euler_plate_lattice}, then the following transformations also yield solutions:
	\begin{enumerate}[label=\textit{(E.\arabic*)}]
		\item (Time translations)\label{SYM:time_bound}
		$
		\!
		\begin{aligned}[t]
			\mathbf{u}^\tau(\mathbf{k},t) = \mathbf{u}(\mathbf{k},t+\tau),
		\end{aligned}
		$
		for any $\tau \in \mathbb{R}$;
		\item (Space translations in $x$ and $z$)\label{SYM:space_bound}
		$
		\!
		\begin{aligned}[t]
			\mathbf{u}^{\pmb{\xi}}(\mathbf{k},t) 
			= e^{-i\mathbf{k}\cdot \pmb{\xi}}\mathbf{u}(\mathbf{k},t),
		\end{aligned}
		$
		for any $\pmb{\xi} = (\xi_x,0,\xi_z) \in \mathbb{R}^3$;
		\item (Isotropy in $x$ and $z$ and parity)\label{SYM:isotropy_bound}
		$
		\mathbf{u}^R(\mathbf{k},t) = R^{-1}\mathbf{u}(R\mathbf{k},t),
		$
		where $R$ is any transformation $(k_1,k_2,k_3) \mapsto (\pm k_\alpha, \pm k_2, \pm k_\beta)$ with $(\alpha,\beta)$ permutations of $(1,3)$;
		\item (Scale invariance)\label{SYM:scaling_bound}
		$
		\!
		\begin{aligned}[t]
			\mathbf{u}^{n,h}(\mathbf{k},t) = \lambda^h \mathbf{u}
			\left(\lambda^{n} \mathbf{k},\lambda^{h-n} t \right),
		\end{aligned}
		$
		for any $h \in \mathbb{R}$ and $n \in \mathbb{Z}$, 
		where $\lambda$ is the lattice spacing;
		\item (Time reversibility)\label{SYM:timerev_bound}
		$
		\!
		\begin{aligned}[t]
			\mathbf{u}^{r}(\mathbf{k},t) = -\mathbf{u}
			\left(\mathbf{k},-t \right)
		\end{aligned}
		$;
		\item (Galilean invariance in $x$ and $z$)\label{SYM:galilean_bound}
		$
		\!
		\begin{aligned}[t]
			\mathbf{u}^{\mathbf{v}}(\mathbf{k},t) = e^{-i\mathbf{k}\cdot \mathbf{v}t}
			\mathbf{u}(\mathbf{k},t) - \widehat{\mathbf{v}}(\mathbf{k}),
		\end{aligned}
		$
		for any $\mathbf{v} = (v_x,0,v_z) \in \mathbb{R}^3$, where $\widehat{\mathbf{v}}(\mathbf{k})$ is the constant velocity field on the lattice defined as $\widehat{\mathbf{v}}(\mathbf{0}) = \mathbf{v}$ and zero for $\mathbf{k} \neq \mathbf{0}$.
	\end{enumerate}
	Naturally, Galilean invariance~\ref{SYM:galilean_bound} is well-defined only if we add the zero component to the lattice $0 \in \mathbb{\Lambda}$.
	
	Since the equations are not modified by introducing a boundary, the conserved quantities are also the same, say: the energy
	\begin{equation}\label{EQ:euler_energy}
		E(t) = \frac{1}{2}\sum_{\mathbf{k} \in \mathbb{\Lambda}^d}|\mathbf{u}(\mathbf{k},t)|^2
	\end{equation}
	and helicity
	\begin{equation}\label{EQ:euler_helicity}
		H(t) = \sum_{\mathbf{k} \in \mathbb{\Lambda}^d}\mathbf{u}(\mathbf{k},t)\overline{\pmb{\omega}(\mathbf{k},t)}
	\end{equation}
	in the three-dimensional case $d=3$, and the energy~\eqref{EQ:euler_energy} and enstrophy
	\begin{equation}\label{EQ:euler_enstrophy}
		\Omega(t) = \frac{1}{2}\sum_{\mathbf{k} \in \mathbb{\Lambda}^d}|\pmb{\omega}(\mathbf{k},t)|^2
	\end{equation}
	in the two-dimensional case $d =2$.
	Here, $\pmb{\omega} = \nabla \times \mathbf{u}$ are the vorticities.
	Kelvin's Theorem for the conservation of circulation also holds in the presence of a boundary.
	The precise statements and the proofs for those conservations laws can be found in~\cite{campolina2019fluid,campolina2021fluid}.
	
	In the case of positive viscosity, the Navier-Stokes equations with and without boundary share the same scaling symmetry
	\begin{enumerate}[label=\textit{(NS)}]
		\item (Scale invariance)\label{SYM:NSscaling_bound}
		$
		\!
		\begin{aligned}[t]
			\mathbf{u}^{n}(\mathbf{k},t) = \lambda^{-n} \mathbf{u}
			\left(\lambda^{n} \mathbf{k},\lambda^{-2n} t \right),
		\end{aligned}
		$
		for any $n \in \mathbb{Z}$, 
		where $\lambda$ is the lattice spacing.
	\end{enumerate}
	Thus, the introduction of a boundary through our modelling technique does not disrupt the self-similarity properties of the Navier-Stokes equations.
	
	Additionally, the shear force~$\mathbf{F}(k_x,k_y,k_z,t) = -\nu \mathbf{J}(k_x,k_z,t)\delta(k_y)$ on the boundary exerts no work in the flow, as a consequence of no-slip boundary condition
	\begin{equation}
		(\mathbf{F},\mathbf{u}) = -\nu \sum_{k_x,k_z \in  \mathbb{\Lambda}}\mathbf{J}(k_x,k_z,t) \cdot (\mathbf{u},\delta)_{k_y} = 0.
	\end{equation}
	This proportionates the classical energy balance law
	\begin{equation}
		\frac{dE}{dt} = -2\nu \Omega(t),
	\end{equation}
	where $\Omega$ is the enstrophy~\eqref{EQ:euler_enstrophy}.
	
	The conservation laws and the energy balance are also satisfied by the truncated flows $\mathbf{u}^N$.
	
	
	\section{Classical shear flows}\label{SEC:1Dshear}
	
	In this section, we take one step back by considering some classical shear flows.
	Due to their simplicity, we can study the logarithmic lattice solutions in the light of exact expected results, or even compare them with direct numerical simulations.
	Naturally, the comparisons are always in terms of qualitative behavior.
	
	Consider the governing system of equations~\eqref{EQ:3D_flow} for a three-dimensional flow on the upper-half space $y>0$ with a solid boundary on the plane $y = 0$.
	Let us assume such flow has no variation with respect to $x$ and $z$, and that $v = w = 0$.
	Under those hypotheses, incompressibility is trivially satisfied, while pressure is constant.
	Then, the resulting flow simplifies to a one-dimensional velocity field $u = u(y,t)$ governed by
	\begin{equation}
		\begin{cases}
			\partial_t u = \nu \partial_y^2 u + f\quad &\text{in} \ y>0,\\
			u = 0 \quad &\text{on} \ y = 0,
		\end{cases}
		\label{EQ:simpler_system}
	\end{equation}
	where $f = f(y,t)$ is a possible external force.
	System~\eqref{EQ:simpler_system} is supplemented by proper initial conditions
	\begin{equation}
		u\Big|_{t=0} = u^0(y).
	\end{equation}
	
	This is the Dirichlet problem for the one-dimensional linear heat equation.
	Such system can be solved exactly using the heat kernel and reflections---see Appendix~\hyperref[app:B]{B} for a closed formula.
	Here, we understand its solutions as simple shear flows over a solid plate at $y=0$.
	
	
	
	Following the steps from Section~\ref{SEC:solid_singular}, we can formulate problem~\eqref{EQ:simpler_system} on the whole domain, considering a discontinuity point at the origin $y = 0$.
	The resulting immersed boundary formulation reads
	\begin{equation}
		\begin{cases}
			\partial_t u = \nu \partial_y^2 u  + f - \nu J(t)\delta(y) \quad &\text{in} \ \mathbb{R},\\
			u(y) = u(-y) \quad &\text{in} \ \mathbb{R},\\
			u = 0 \quad &\text{on} \ y = 0.
		\end{cases}
		\label{EQ:simpler_system_control}
	\end{equation}
	The jump discontinuity is derived from system~\eqref{EQ:simpler_system_control} as
	\begin{equation}
		J(t) = [\partial_y u].
	\end{equation}
	
	On a logarithmic lattice
	\begin{equation}\label{EQ:lattice_full}
		\mathbb{\Lambda} = \{ 0, \pm 1, \pm \lambda, \pm \lambda^2, \dots \},
	\end{equation}
	model~\eqref{EQ:simpler_system_control} reads
	\begin{equation}
		\begin{cases}
			\partial_t u = \nu \partial_y^2 u  + f -\nu J(t)\delta(k) \quad &\text{in} \ \mathbb{\Lambda},\\
			u(k) = u(-k) \quad &\text{in} \ \mathbb{\Lambda},\\
			(u,\delta) = 0 \quad &\text{for all} \ t \in \mathbb{R}.
		\end{cases}
		\label{EQ:simpler_system_control_lattice}
	\end{equation}
	
	As we saw earlier, the jump $J(t)$ can be computed explicitly.
	If the lattice~\eqref{EQ:lattice_full} is truncated up to the wave number $k_N = \lambda^{N-1}$ as
	\begin{equation}\label{EQ:truncated_lattice_1Dshear}
		\mathbb{\Lambda}_N = \{ 0, \pm 1, \pm \lambda, \pm \lambda^2, \dots \lambda^{N-1} \},
	\end{equation}
	we have the truncated model
	\begin{equation}
		\begin{cases}
			\partial_t u^N = \nu \partial_y^2 u^N + f^N -\nu J^N(t)\delta^N(k) \quad &\text{in} \ \mathbb{\Lambda}_N,\\
			u^N(k) = u^N(-k) \quad &\text{in} \ \mathbb{\Lambda}_N,\\
			(u^N,\delta^N) = 0 \quad &\text{for all} \ t \in \mathbb{R}.
		\end{cases}
		\label{EQ:simpler_system_control_lattice_N}
	\end{equation}
	
	Taking the inner product of the main equation in~\eqref{EQ:simpler_system_control_lattice_N} against $\delta^N$, and using the no-slip condition $(u^N,\delta^N) = 0$, we obtain
	\begin{equation}\label{EQ:partial_J}
		(\nu \partial_y^2 u^N + f^N, \delta^N) + (-\nu J^N(t)\delta^N,\delta^N) = 0.
	\end{equation}
	Next, using the fact that $J^N(t)$ does not depend on $k$, we can evaluate the term
	\begin{equation}\label{EQ:expression_for_F}
		(-\nu J^N(t)\delta^N,\delta^N) = -\nu J^N(t) (\delta^N,\delta^N).
	\end{equation}
	Observe that, since the lattice is now truncated, the expression $(\delta^N,\delta^N)$ is the well-defined positive number
	\begin{equation}
		(\delta^N,\delta^N) = \sum_{k \in \mathbb{\Lambda}^N}1
		= 2N + 1.
	\end{equation}
	Finally, we substitute~\eqref{EQ:expression_for_F} into~\eqref{EQ:partial_J} and isolate $J^N(t)$ to obtain a closed formula for the approximated jump
	\begin{equation}
		J^N(t) = \frac{(\partial_y^2 u^N + f^N,\delta^N)}{(\delta^N,\delta^N)}.
	\end{equation}
	
	The original jump is then recovered from the sequence of its approximations
	\begin{equation}
		J(t) = \lim\limits_{N \to \infty} J^N(t)
		= \lim\limits_{N \to \infty} \frac{(\partial_y^2 u^N + f^N,\delta^N)}{(\delta^N,\delta^N)}.
	\end{equation}
	
	\subsection{Couette flow}\label{SEC:pipe}
	
	Let us consider a flow between two parallel plates separated by a unit distance.
	One of the plates is at rest, and the other moves with a constant horizontal speed $V$.
	This classical problem has well-known stationary solution given by the linear velocity profile
	\begin{equation}\label{EQ:shear_profile}
		u(y) = Vy \quad \text{for} \ 0 \leq y \leq 1.
	\end{equation}
	Such configuration is called \textit{Couette flow}---consult \textit{e.g.}~\cite[\S 17]{landau1987fluid}.
	Then, the shear force~$f_1$ exerted on the fluid by the moving plate at $y = 1$ is
	\begin{equation}\label{EQ:force_original_parallel_1}
		f_1 = \nu V \quad \text{at} \ y =1,
	\end{equation}
	while the force~$f_0$ from the plate~$y = 0$ at rest is the symmetric counterpart
	\begin{equation}\label{EQ:force_original_parallel_0}
		f_0 = -\nu V \quad \text{at} \ y =0,
	\end{equation}
	
	To model a similar phenomenon on a logarithmic lattice, we take the one-dimensional shear flow equation~\eqref{EQ:simpler_system_control_lattice} and consider the action of the moving plate as a constant-in-time force $f$ applied at $k = 0$ in the form
	\begin{equation}\label{EQ:force_V}
		f(k,t) = 2\nu V \delta_0(k),
	\end{equation}
	with
	\begin{equation}
		\delta_0 (k) = 
		\begin{cases}
			1, \quad \text{if} \ k = 0,\\
			0, \quad \text{otherwise}.
		\end{cases}
	\end{equation}
	
	We concentrate the force at~$k = 0$ in order to model the momentum input due to the moving boundary.
	The choice of the force~\eqref{EQ:force_V} is motivated by the known shear action~\eqref{EQ:force_original_parallel_1} at the moving plate from the original problem.
	Such force is proportional to the relative velocitity between the plates and to the fluid viscosity.
	The factor $2$ appears in~\eqref{EQ:force_V} because our reflected flow doubles the forces on the discontinuity surface.
	
	\subsubsection{Stationary solution}
	
	Let us look for stationary solutions of system~\eqref{EQ:simpler_system_control_lattice} under the action of the force~\eqref{EQ:force_V}.
	Evaluating the governing equation at $k = 0$ gives us the value of the jump
	\begin{equation}
		J = 2V.
	\end{equation}
	The solution at $k \neq 0$ can now be obtained, using that $f(k) = 0$ at $k \neq 0$, as
	\begin{equation}
		0 = \nu\partial^2_yu(k) - J\delta(k)
		= -\nu k^2 u(k) - \nu J,
	\end{equation}
	which yields
	\begin{equation}
		u(k) = -2Vk^{-2} \quad \text{for} \ k \neq 0.
	\end{equation}
	Finally, the mean flow $u(k = 0)$ can be computed from the no-slip boundary condition, as follows
	\begin{equation}
		0 = (u,\delta) = \sum_{k \in \mathbb{\Lambda}}u(k)
		= u(0) + \sum_{k \in  \mathbb{\Lambda} \backslash \{0\}}u(k)
		= u(0) - 2V\sum_{k \in  \mathbb{\Lambda} \backslash \{0\}}k^{-2},
	\end{equation}
	whence
	\begin{equation}
		u(0) = 2V\sum_{k \in  \mathbb{\Lambda} \backslash \{0\}}k^{-2}.
	\end{equation}
	
	The final solution
	\begin{equation}\label{EQ:stationary_shear}
		J = 2V \quad \text{and} \quad u(k) =
		\begin{cases}
			2V\sum_{\tilde{k} \in  \mathbb{\Lambda} \backslash \{0\}}\tilde{k}^{-2} \quad &\text{for} \ k = 0,\\
			-2Vk^{-2} \quad &\text{for} \ k \neq 0,
		\end{cases}
	\end{equation}
	shares many similitudes with the original shear flow.
	First, the jump agrees exactly with the original velocity profile~\eqref{EQ:shear_profile}, if we consider its symmetric reflection $u(-y) = u(y)$ around the origin $y = 0$.
	Consequently, the shear force at the plate is~$-\nu J = -2\nu V$, which is in agreement with the force~\eqref{EQ:force_original_parallel_0}.
	Remember that the reflections of the flow around the origin double the forces, which explains the extra factor $2$.
	Second, we observe a solution tail proportional to $k^{-2}$, which is the expected Fourier spectrum for a function whose first derivative is discontinuous~\cite{friedlander1998introduction}.
	
	\subsubsection{Unsteady solutions}
	
	We can also simulate unsteady solutions of system~\eqref{EQ:simpler_system_control_lattice} under the action of the force~\eqref{EQ:force_V}.
	For this, we set $V = 1$, $\nu = 1$ and consider identically zero initial conditions $u^0 \equiv 0$.
	We take the truncated logarithmic lattice~\eqref{EQ:truncated_lattice_1Dshear} with $\lambda = 2$ and $N = 50$.
	Together, they provide the finest scale $\ell_N = 1/k_N \approx 10^{-15}$.
	We integrate the equations using \textsc{Matlab}'s ode15s solver~\cite{shampine1997matlab}, with the tolerances $RelTol = 10^{-8}$ and $AbsTol = 10^{-11}$.
	
\begin{figure*}[t]
		\centering
		\includegraphics[width=\textwidth]{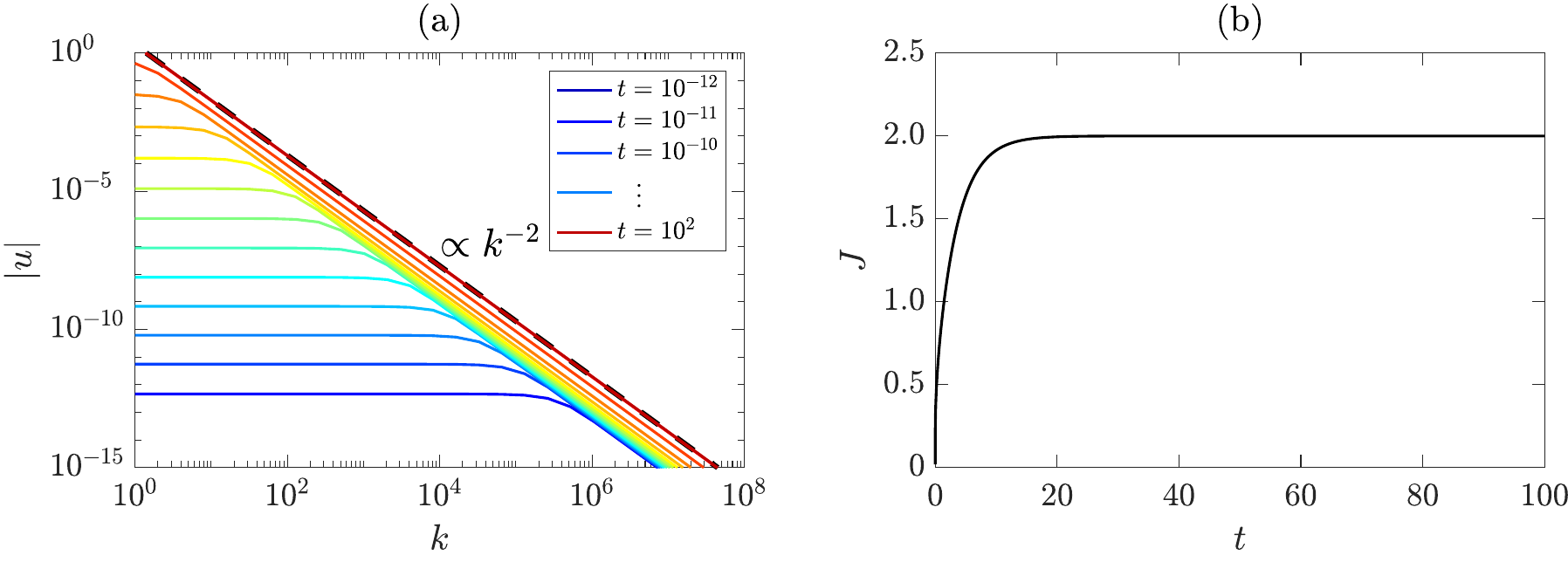}
		\caption{Dynamical evolution of the Couette flow on a logarithmic lattice.
			(a)~Spectra of solution $|u(k)|$ as a function of the wave number $k$ at several instants.
			Time advances from blue to red in logarithmic scale, from $t = 10^{-12}$ up to $t = 10^{2}$ by factors of $10$.
			The solution converges to the stationary power-law $k^{-2}$, represented by the black dashed line.
			(b)~Time evolution of the jump variable $J(t)$, converging to the constant steady value $J = 2V$, with $V=1$, as time advances.}
		\label{FIG:shearV1}
	\end{figure*}

	Fig.~\ref{FIG:shearV1} shows the time evolution of our Couette-like flow.
	As time advances, the solution converges to the stationary state~\eqref{EQ:stationary_shear}, which is a fixed point attractor of the system.
	At each instant, the solution presents two different regimes along scales: a constant plateau $|u| \approx const.$ state at larger scales, followed by a $|u| \propto k^{-2}$ tail at smaller scales.
	Such behavior is more noticeable in earlier instants, since the plateau range shrinks as time advances.
	These two regimes can be understood by establishing the corresponding asymptotic solutions, as we do now.
	For this analysis, we consider that the lattice $\mathbb{\Lambda}$ has infinitely large and infinitely small $k$.
	
	The asymptotic solutions are obtained by analysing how the terms in the governing equation scale with respect to $k$.
	First, we notice that the boundary force~$-\nu J(t)\delta(k)$ acts with the same magnitude at all scales, since $\delta \equiv 1$.
	Therefore, it contributes both to large and to small scales.
	Next, we analyze the contribution of the dissipative term $\nu\partial_y^2 u(k,t) = -\nu k^2 u(k,t)$, which determines the two different regimes.
	At high $k$, the dissipative term becomes important and dominates over the time derivative $\partial_t u$.
	In this regime, we have the balance
	\begin{equation}
		\nu \partial_y^2 u \sim \nu J\delta \quad \text{as} \ k \to \infty,
	\end{equation}
	which results in the tail asymptotic solution
	\begin{equation}\label{EQ:asymptotic_high_k}
		u(k,t) \sim -J(t)k^{-2} \quad \text{as} \ k \to \infty.
	\end{equation}
	
	\begin{figure*}[t]
		\centering
		\includegraphics[width=.6\textwidth]{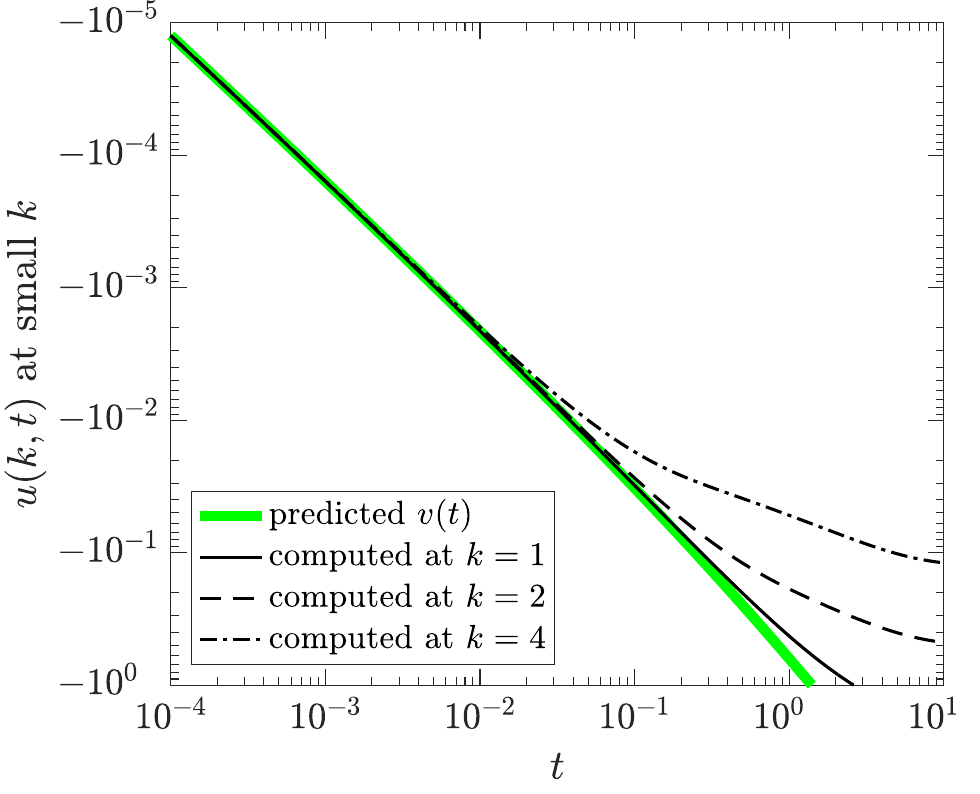}
		\caption{Time evolution of $u(k,t)$ at small $k \in \mathbb{\Lambda}$ on the logarithmic lattice.
			Axes are in logarithmic scales.
			We plot in green the predicted asymptotic value~$v(t)$.
			Black curves show actual computed solutions $u(k,t)$.
			Different styles of lines stand for different scales $k$.}
		\label{FIG:shearV2}
	\end{figure*}
	
	On the other hand, if $k$ is small, the dissipative term becomes negligible, and it is dominated by the time derivative $\partial_t u$.
	The balance in this case is
	\begin{equation}\label{EQ:asymptotic_small_k}
		\partial_t u \sim -\nu J\delta \quad \text{as} \ k \to 0.
	\end{equation}
	Eq.~\eqref{EQ:asymptotic_small_k} is independent from $k$, which justifies the constant plateau behavior
	\begin{equation}\label{EQ:u_plateau}
		u(k,t) \approx v(t) \quad \text{for small} \ k,
	\end{equation}
	in Fig.~\ref{FIG:shearV1}(a).
	To confirm such statement, let us consider the ordinary differential equation
	\begin{equation}\label{EQ:approximated_ODE}
		\frac{dv}{dt} = -\nu J(t),
	\end{equation}
	which is the exact form for the asymptotic balances~\eqref{EQ:asymptotic_small_k} and~\eqref{EQ:u_plateau}.
	We integrate this equation numerically using the numerical result of $J(t)$ from our model.
	Fig.~\ref{FIG:shearV2} compares this predicted value for the plateau with the actual solution $u(k,t)$ for small $k$.
	We verify that as $k$ decreases, the agreement between the prediction and the solution improves.
	
	We observe the transition between the two regimes when $|u(k,t)|$ is of order of $k^2$, which eventually occurs at any finite scale $k$ after sufficiently large time.
	We can see the transition of regimes in Fig.~\ref{FIG:shearV1}(a) and the expected deviation from the asymptotic solution at finite $k$ after some time in Fig.~\ref{FIG:shearV2}.
	
	\subsection{Decaying shear flow}\label{SEC:decaying}
	
	Next we consider a decaying shear flow.
	The setup consists of system~\eqref{EQ:simpler_system} with zero external force $f \equiv 0$ and non-zero initial condition $u^0 \not\equiv 0$.
	
	First, we shall solve the continuous-space immersed boundary formulation~\eqref{EQ:simpler_system_control} with direct numerical simulations.
	For this, we employ simple finite difference schemes with a regularized Dirac delta parametrized by the number of grid points.
	Such strategy will illustrate the explicit computation of jump singularities.
	Then, we shall consider the logarithmic lattice formulation.
	The results on logarithmic lattices will be compared with the precedent direct numerical simulations.
	
	\subsubsection{Direct numerical simulations in continuous space}\label{SEC:decaying_DNS}
	
	Consider the one-dimensional shear flow equation with immersed boundary~\eqref{EQ:simpler_system_control} in continuous space $y \in \mathbb{R}$, with $f \equiv 0$.
	For the numerical model, we shall restrict the domain to the bounded interval $y \in [-L,L]$ and simulate the dynamics on a finite interval of time $t \in [0,T]$.
	Such framework is a good approximation of the unbounded system $y \in \mathbb{R}$ when the initial condition has fast decrease at infinity and $T$ is sufficiently small.
	In what follows, we fix $L = 10$.
	
	We discretize both space and time and consider second-order centered finite differences for the laplacian and first-order forward finite difference for the time derivative~\cite{leveque2007finite}.
	The jump can be explicitly computed with the discrete variables.
	This yields a recurrence relation that can be used to have numerical approximations of the dynamics.

	\begin{figure*}[t]
		\centering
		\includegraphics[width=\textwidth]{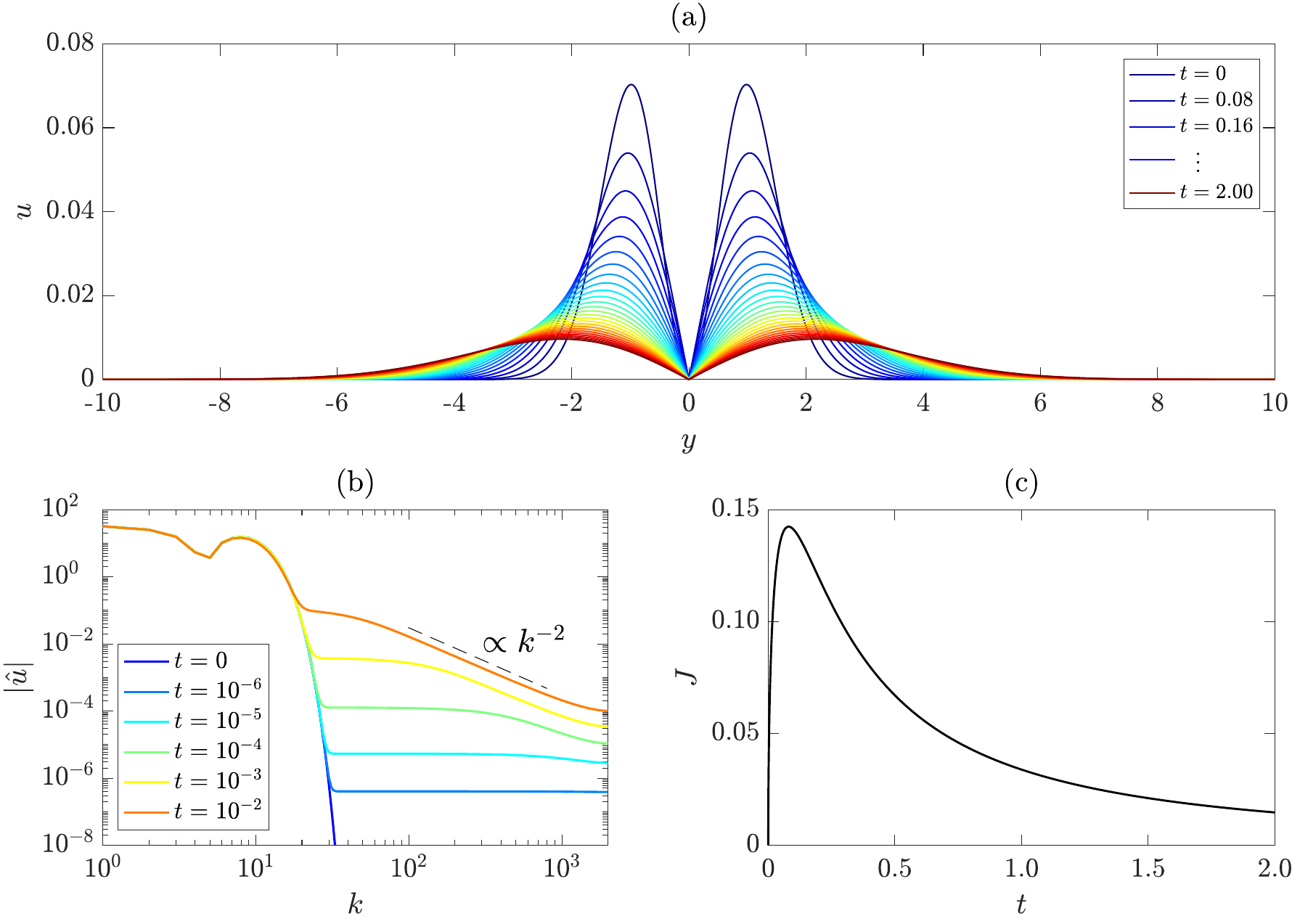}
		\caption{Direct numerical simulations of the continuous-space immersed boundary formulation of a decaying shear flow. (a)~Solution $u$ along space $y$.
			Different colors indicate solutions at different instants $t$.
			Time advances from blue to red.
			(b)~Fourier spectra of solution in logarithmic scales.
			Time advances from from blue to red in logarithmic scale.
			(c)~Time evolution of the jump $J(t)$.}
		\label{FIG:solution}
	\end{figure*}
	
%
%
	
	For the numerical experiments, we set $\nu = 1$, and the initial condition to
	\begin{equation}\label{EQ:IC_finite_difference}
		u^0(y) = \left[ 1 - \cos \left( \frac{\pi y}{5} \right) \right] e^{-y^2/2} \quad \text{for} \ y \in \mathbb{R}.
	\end{equation}
	This function satisfies the odd symmetry $u(-y) = u(y)$ and decays exponentially fast at infinity $|y| \to \infty$.
	Moreover, the initial condition satisfies the no-slip condition $u^0(0) = 0$ and has zero initial shear force as $[\partial_yu^0] = 0$.
	
	Our first numerical integrations concern the general picture of the solution in physical space after some amount of time.
	We use $T = 2$, $N = 1024$ points in space and $M = 20~000$ points in time.
	The results are depicted in Figs.~\ref{FIG:solution}(a) and~\ref{FIG:solution}(c).
	To reach better resolution in Fourier space at earlier instants, we change the numerical parameters to $T = 0.01$ and $N = 4096$, while $M$ is kept the same.
	The Fourier spectrum is drawn in Fig.~\ref{FIG:solution}(b).
	We now discuss the results.
	
	We show the time evolution of the velocity along physical space in Fig.~\ref{FIG:solution}(a).
	The initial data~\eqref{EQ:IC_finite_difference} is essentially two symmetric bumps close to the origin quickly decaying at infinity.
	They represent an initial motion restricted to large scales---as we verify from the spectrum concentrated on small $k$ in Fig.~\ref{FIG:solution}(b).
	As time advances, the solution dissipates and tends towards the identically zero steady state.
	Initially, the velocity profile develops a large increase in the jump, it achieves a maximum and then decays, as in Fig.~\ref{FIG:solution}(c).
	
	In Fourier space, we see some features that we already presented for the logarithmic model of a Couette-like flow.
	The initial state is confined to large scales, but the boundary shear force affects all scales at every instant of time $t>0$.
	This is noticed by the instantaneous development of a constant plateau regime along intermediate scales.
	Such plateau would be followed by a $k^{-2}$ tail at large scales even for very small $t$, if we simulated for larger $k$.
	Under limited resolution, we can see the power-law development at later times in Fig.~\ref{FIG:solution}(b), when the tail already reaches intermediate and large scales.
	We observe some slight deviation from the $k^{-2}$ at the higher scales, as an effect of truncation.
	Increasing resolution extends the power-law and sends this truncation effect towards higher $k$.
	
	\subsubsection{Logarithmic lattice simulations}\label{SEC:decaying_loglattice}
	
	\begin{figure*}[t]
		\centering
		\includegraphics[width=\textwidth]{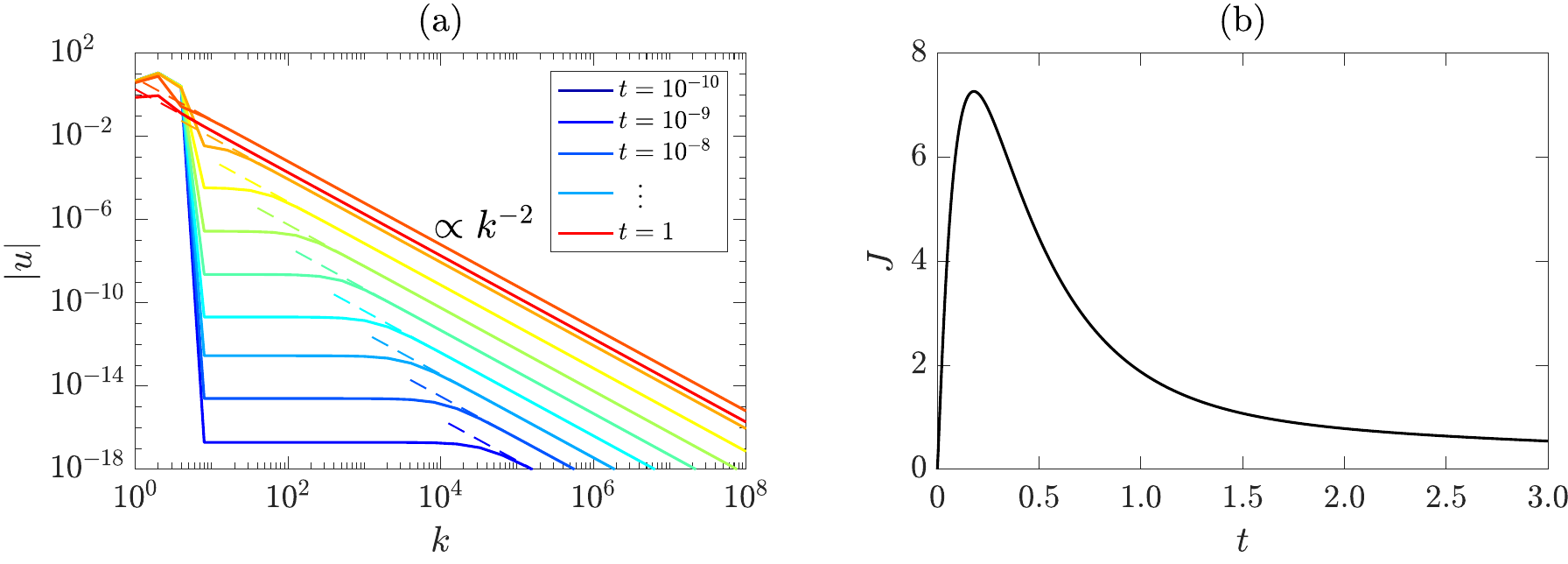}
		\caption{Dynamical evolution of a decaying shear flow on a logarithmic lattice.
			(a)~Spectra of solution $|u(k)|$ as a function of the wave number $k$ at several instants.
			Time advances from blue to red in logarithmic scale, from $t = 10^{-10}$ up to $t = 10^{0}$ by factors of $10$.
			The dashed lines show the corresponding asymptotic solutions $u(k,t) \sim -J(t)k^{-2}$ for large $k$.
			(b)~Time evolution of the jump variable $J(t)$.}
		\label{FIG:shear_decay}
	\end{figure*}
	
	Now, let us consider the decaying shear flow on a logarithmic lattice.
	We take the governing system of equations~\eqref{EQ:simpler_system_control_lattice} with unit viscosity $\nu = 1$ and zero external force $f \equiv 0$.
	Representing the initial condition $u(k,0)$ by the vector $u^0 = (u^0(0), u^0(\pm1), u^0(\pm\lambda), u^0(\pm\lambda^2), \dots)$, we fix
	\begin{equation}
		u^0 = \left( 8, \frac{14}{3}, -\frac{67}{6}, \frac{5}{2}, 0, 0, \dots \right).
	\end{equation}
	Such initial flow satisfies the no-slip condition $(u^0,\delta) = 0$ and has no initial shear flow at the boundary, since $(\partial_y^2u^0,\delta) = 0$, and thus $J|_{t = 0} = 0$.
	
	The numerical simulations are undertaken on the truncated lattice~\eqref{EQ:truncated_lattice_1Dshear} with $\lambda = 2$ and $N = 50$.
	We solve the equations with \textsc{Matlab}'s ode15s solver~\cite{shampine1997matlab}, with tolerances $RelTol = 10^{-8}$ and $AbsTol = 10^{-11}$.
	
	\begin{figure*}[t]
		\centering
		\includegraphics[width=.85\textwidth]{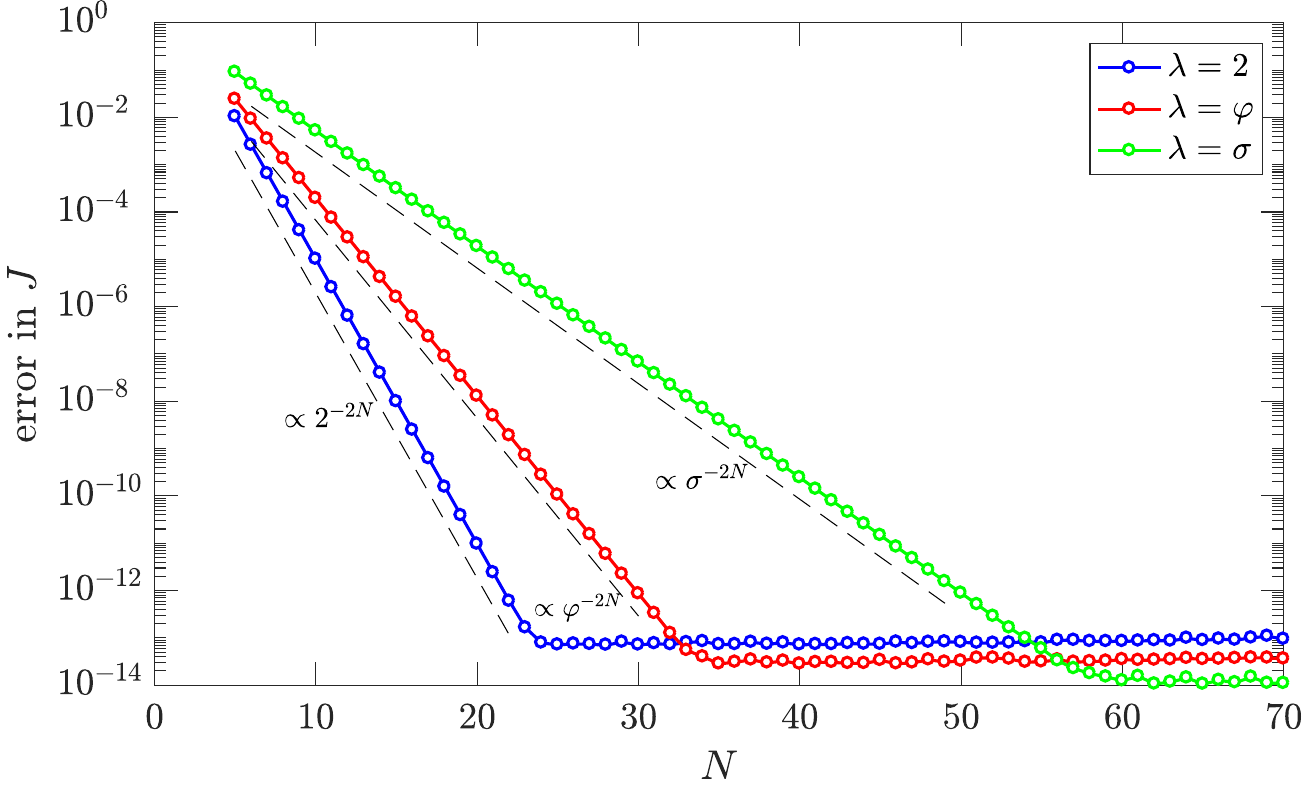}
		\caption{Convergence of solutions with respect to truncation.
			Error in $\ell^2$ norm in time for the jump $J^N$ with respect to a reference solution with $75$ as a function of the number of nodes $N$.
			Different curves stand for the three lattice spacings: $\lambda = 2$ in blue, $\lambda = \varphi$ (the golden mean) in red, and $\lambda = \sigma$ (the plastic number) in green.
			Dashed lines give reference to exponential convergence $\propto \lambda^{-2N}$.}
		\label{FIG:shear_convergence}
	\end{figure*}
	
	In Fig.~\ref{FIG:shear_decay} we see the dynamics of the solution on the logarithmic lattice.
	Fig.~\ref{FIG:shear_decay}(a) shows the spectra of the velocites for different instants of time.
	The initial spectrum is confined to large scales, but the flow instantly develops a $k^{-2}$ tail at high $k$.
	The gap between large and small scales is filled by a constant plateau state, which shrinks as time advances.
	In dashed lines, we plot the asymptotic solutions~\eqref{EQ:asymptotic_high_k} and verify that they match the computed results at high $k$.
	
	Here we verify that the solution on the logarithmic lattice is quite similar to that obtained from DNS, with the advantage that in the former we can reach way finer scales.
	The observations of a $k^{-2}$ tail, a plateau in intermediate scales and the dynamical shrink of the gap between them is also present in Fig.~\ref{FIG:solution}(b) for the continuous space model, but in a more restricted resolution.
	Moreover, the time evolution of the jump lattice variable shown in Fig.~\ref{FIG:shear_decay}(b) is qualitatively indistinguishable from the corresponding jump on the continuous model from DNS in Fig.~\ref{FIG:solution}(c).
	
	\subsection{Convergence with respect to truncation}\label{SEC:convergence_truncation}
	
	Heuristically, truncation of the logarithmic lattice $\mathbb{\Lambda}_N$ at a sufficiently large wave number $k_N = \lambda^{N-1}$ represents a cutoff of the $k^{-2}$ tail whose sum is proportional to $k_N^{-2}$.
	This establishes the exponential convergence rate $\lambda^{-2N}$.
	
	To verify such convergence rate with respect to truncation, we perform simulations of the decaying shear flow from Section~\ref{SEC:decaying_loglattice} for several number of nodes.
	We range $N$ from $5$ up to $70$ by unit increments.
	To compute errors, we consider the simulation $N = 75$ as reference.
	We integrate equations using \textsc{Matlab}'s ode15s solver with tolerances $RelTol = 10^{-13}$ and $AbsTol = 10^{-16}$.
	We fix the time window $T = 3$ and sample the solution at points $t_j = j \Delta t$, for $j = 0,1,\dots,M$ with $M = 3000$, so $\Delta t = T/M = 0.001$.
	For each $N$, we compute the $\ell^2$ norm in time of the difference between the jump $J^N$ with respect to the jump of the reference solution with $75$ nodes.
	
	Fig.~\ref{FIG:shear_convergence} shows the results of this set of runs.
	We readily attest the exponential rate of convergence $(error)_N \propto \lambda^{-2N}$ for the main lattice spacings $\lambda = 2$, $\varphi$ (the golden ratio $\varphi \approx 1.618$) and $\sigma$ (the plastic number $\sigma \approx 1.325$).
	Since the convergence depends on the lattice spacing, errors reach minimum value at a different number of node points, which is approximately $N = 24$ for $\lambda = 2$, $N = 35$ for $\lambda = \varphi$ and $N = 60$ for $\lambda = \sigma$.
	
	
	\section{Two-dimensional boundary layers}\label{SEC:2DBL}
	
	In this section, we study the development of boundary layers over solid boundaries and the convergence of Navier-Stokes solutions to Euler's at infinite Reynolds limit in this framework.
	Just like in reported DNS~\cite{nguyen2011energy,nguyen2018energy}, we study this problem in two-dimensions.
	In the light of Kato's Equivalence Theorem~\cite{kato1984remarks}, we track total dissipation of Navier-Stokes flows with increasing Reynolds number.
	Here, however, we do not give a rigorous final answer to this problem on logarithmic lattices.
	Many questions remain open and further work is needed.
	The main goal of this chapter is to present promising results for this long standing problem.
	Particularly, we show that logarithmic models can reach extremely larger Reynolds numbers than in DNS.
	
	We consider two-dimensional flow in the presence of a flat boundary at $y = 0$.
	Using the two-dimensional logarithmic lattice $\mathbf{k} = (k_x,k_y) \in \mathbb{\Lambda}^2$,
	the velocity field $\mathbf{u}(\mathbf{k},t) = (u,v)$ and the scalar pressure $p(\mathbf{k},t)$ are governed by the incompressible Navier-Stokes equations with a solid boundary~\eqref{EQ:flat_plate_model_lattice}, and can be explicitly written in the form
	\begin{equation}\label{EQ:two_dimensional_BL_log_lattice}
		\begin{cases}
			\partial_t u + u \ast \partial_x u + v \ast \partial_y u = - \partial_x p + \nu \Delta u -\nu J(k_x,t)\delta(k_y) \quad &\text{in} \ \mathbb{\Lambda}^2 \\
			\partial_t v + u \ast \partial_x v + v \ast \partial_y v = - \partial_y p + \nu \Delta v \quad &\text{in} \ \mathbb{\Lambda}^2 \\
			\partial_x u + \partial_y v = 0 \quad &\text{in} \ \mathbb{\Lambda}^2 \\
			u(-k_y) = u(k_y), \quad v(-k_y) = -v(k_y), \quad p(-k_y) = p(k_y) \quad &\text{in} \ \mathbb{\Lambda}^2 \\
			(u,\delta)_{k_y} = 0 \quad &\text{for all} \ k_x \in \mathbb{\Lambda}.
		\end{cases}
	\end{equation}
	In two dimensions, the boundary shear force acts only in the $x$ direction.
	Therefore we represent the jump variable $J$ as a scalar.
	
	As usual, the pressure can be eliminated from the equation due to incompressibility, by solving the Poisson equation~\eqref{EQ:poisson_pressure_boundary}.
	In the case of non-zero viscosity, the pressure will be solved in terms of velocities and the jump $J$.
	
	\subsubsection*{Computation of the jump}
	
	In the numerical integrations, we consider system~\eqref{EQ:two_dimensional_BL_log_lattice} on the truncated logarithmic lattice $\mathbb{\Lambda}_N^2$, given by
	\begin{equation}
		\mathbb{\Lambda}_N = \{ \pm 1, \pm\lambda, \pm\lambda^2, \dots \pm\lambda^{N-1} \}.
	\end{equation}
	As we did in the previous sections, we add superscript $N$ to the variables to indicate the solutions of the truncated system.
	
	To compute the jump $J^N$, we take the $k_y$-inner product of the $u^N$ equation against $\delta^N$ and use the no-slip condition $(u^N,\delta^N)_{k_y} = 0$ to eliminate the time derivative contribution $(\partial_t u^N,\delta)_{k_y} = 0$.
	After solving for the pressure $p^N$, the approximated jump $J^N$ can be isolated and it has the following explicit formula
	\begin{equation}\label{EQ:jump_two_dimensions}
		J^N(k_x,t) = \frac{1}{\nu (\xi \delta^N,\delta^N)}_{k_y} (\partial_x \Delta^{-1}\nabla \cdot (\mathbf{u}^N \ast \nabla \mathbf{u}^N) - u^N \ast \partial_x u^N - v^N \ast \partial_y u^N + \nu \partial_y^2 u^N, \delta^N)_{k_y},
	\end{equation}
	where $\Delta^{-1}$ and $\xi$ are the Fourier multipliers
	\begin{equation}
		\Delta^{-1}(k_x,k_y) = \frac{1}{k_x^2 + k_y^2}, \quad \xi(k_x,k_y) = \frac{k_y^2}{k_x^2 + k_y^2}.
	\end{equation}
	The jump $J$ is recovered from the approximated jumps $J^N$ throught the limit
	\begin{equation}
		J(k_x,t) = \lim\limits_{N \to \infty} J^N(k_x,t).
	\end{equation}
	
	\subsubsection*{Convergence of jump variable}
	
	Let us analyse expression~\eqref{EQ:jump_two_dimensions} for the truncated jump $J^N$.
	We can write it in terms of two contributions
	\begin{equation}
		J^N = NL^N + Diss^N,
	\end{equation}
	where $NL^N$ represents the contribution of nonlinear terms
	\begin{equation}
		NL^N = \frac{1}{\nu (\xi \delta^N,\delta^N)}_{k_y} (\partial_x \Delta^{-1}\nabla \cdot (\mathbf{u}^N \ast \nabla \mathbf{u}^N) - u^N \ast \partial_x u^N - v^N \ast \partial_y u^N, \delta^N)_{k_y}
	\end{equation}
	and $Diss^N$, the contribution of the dissipative term
	\begin{equation}
		Diss^N = \frac{1}{(\xi \delta^N,\delta^N)}_{k_y} (\partial_y^2 u^N, \delta^N)_{k_y}.
	\end{equation}
	If the velocities remain in the same regularity class (with the expected $k^{-2}$ tail in the spectrum), the numerator of $NL^N$ is going to converge to a finite value in the limit $N \to \infty$.
	The denominator, however, grows with respect to $N$, for a fixed viscosity.
	Therefore, we must have
	\begin{equation}
		\lim\limits_{N \to \infty} NL^N = 0,
	\end{equation}
	and only $Diss^N$ effectively contributes to the jump in the limit $N \to \infty$.
	The term $NL^N$ cannot be dropped, however, otherwise the no-slip boundary condition is not satisfied.
	
	\begin{table}[t]
		\centering
		\begin{tabular}{c | c c c c c c c c c c} 
			\hline\hline 
			\textbf{Run}&\textbf{I}&\textbf{II}&\textbf{III}&\textbf{IV}&\textbf{V}&\textbf{VI}&\textbf{VII}&\textbf{VIII}&\textbf{IX}&\textbf{X}\\
			\hline
			$\mathbf{Re}$&$10^{1}$&$10^{2}$&$10^{3}$&$10^{4}$&$10^{5}$&$10^{6}$&$10^{7}$&$10^{8}$&$10^{9}$&$10^{10}$ \\
			$\mathbf{N}$&$60$&$68$&$75$&$83$&$90$&$98$&$105$&$113$&$120$&$128$ \\
			\hline\hline
		\end{tabular}
		\caption{Reynolds number~$Re$ and number of nodes~$N$ for each run.}
		\label{TAB:numerical_experiments}
	\end{table}
	
	A more subtle convergence problem arises in the study of vanishing viscosity.
	A small viscosity parameter appearing in the denominator of $NL^N$ may amplify the spurious contribution of the nonlinearities to the computation of jumps.
	To guarantee convergence, we must first take the truncation limit $N \to \infty$, and later the vanishing viscosity limit $\nu \to 0$.
	Still, for proper numerical simulations, we must have the compensation of the two terms $\nu$ and $(\xi \delta^N,\delta^N)$, in such a way that their product is big enough to provide a small $NL^N$.
	In order to guarantee an adequate convergence rate to $NL^N \to 0$, we redefine the inner product as
	\begin{equation}\label{EQ:inner_alpha}
		(f,g) = \sum_{\mathbf{k} \in  \mathbb{\Lambda}^d}|k_1\dots k_d|^\alpha f(\mathbf{k})\overline{g(\mathbf{k})}
	\end{equation}
	and the star product accordingly
	\begin{equation}\label{EQ:product_beta}
		(f\ast g)(\mathbf{k}) = |k_1 \dots k_d|^\beta \sum_{\substack{\mathbf{p} + \mathbf{q} = \mathbf{k}\\[2pt] \mathbf{p},\mathbf{q} \in \mathbb{\Lambda}^d}} \left|\frac{p_1 \dots p_d}{k_1 \dots k_d}\frac{q_1 \dots q_d}{k_1 \dots k_d}\right|^{\frac{\alpha + \beta}{3}} f(\mathbf{p})g(\mathbf{q}).
	\end{equation}
	The factors $|k_1\dots k_d|$ are interpreted as the volume of lattice cells.
	Parameters $\alpha$ and $\beta$ can be manipulated to change dimensionality and the scaling of terms.
	With such modifications, we do not change the structure of the equations, nor their properties (symmetry group, conservation laws, etc).
	Consult~\cite[\S4]{campolina2022fluid} for more details.
	
	For the generalized inner product~\eqref{EQ:inner_alpha} and the generalized convolution product~\eqref{EQ:product_beta} with parameters $\alpha > 0$ and $\beta = 0$, one may estimate the lower bound
	\begin{equation}
		(\xi \delta^N,\delta^N) \geq \lambda^{\alpha (N-1)} \quad \text{for all} \ k_x \in \mathbb{\Lambda}_N.
	\end{equation}
	Since $(\xi \delta^N,\delta^N)$ grows exponentially fast, the decrease in one order of magnitude in $\nu$ is roughly compensated by a linear increase in $N$.
	In analogy to the Boundary Layer Theory~\cite{schlichting1961boundary}, we expect the jumps $J$ to increase as the viscosity vanishes.
	For this reason, in practice we do not try to keep $NL^N$ small in absolute value, but in relative error with respect to the total jump, \textit{i.e.} we keep $NL^N/J^N$ small.
	
	\subsection{Numerical setup}
	
	We chose the logarithmic lattice with golden ratio spacing $\lambda \approx 1.618$.
	Model~\eqref{EQ:two_dimensional_BL_log_lattice} is integrated with double precision by \textsc{Matlab}'s ode15s solver~\cite{shampine1997matlab}.
	We set the tolerances $RelTol = 10^{-10}$ and $AbsTol = 10^{-13}$.
	For the operations on logarithmic lattices, we employ the computational library \textsc{LogLatt}~\cite{campolina2020loglatt,campolina2020loglattmatlab}.
	
	The initial condition is fixed for all simulations and taken as random components in a box of three by three nodes, at the large scales $(k_x,k_y)$ with $1 \leq k_x,k_y \leq \lambda^2$, and zero elsewhere.
	The components are adjusted to match no-slip condition and zero initial jump.
	
	We simulate Reynolds numbers $Re$ from $10^1$ up to $10^{10}$.
	Parameter $\alpha$ and the number of nodes $N$ in each direction of the lattice are chosen so the relative error $\varepsilon^N = NL^N/J^N$ is kept small.
	We set $\alpha = 0.2$.
	The values of $Re$ and $N$ for each run are in Tab.~\ref{TAB:numerical_experiments}.
	The error $\varepsilon^N$ oscillates along scales due to nonlinearities, but they decrease in average from $10^{-2}$ to $10^{-4}$ as $Re$ increases.
	To ensure that the qualitative behavior do not change because of the error on the jump computation, we performed simulations with twenty more and twenty less nodes.
	All simulations presented the same qualitative behavior, with also good quantitative agreement before the transition to chaos.
	Moreover, the error in $\ell^\infty$ norm for the incompressibility was kept below $8 \times 10^{-15}$ for all simulations at all instants,
	and the error in $\ell^\infty_t \ell^1_{k_x}$ for the slip at the boundary was kept below $1.2 \times 10^{-9}$.
	
	We now turn to the detailed presentation of the results.
	
	\subsection{Laminar to turbulent transition}
	
\begin{figure*}[p]
		\centering
		\includegraphics[width=.7\textwidth]{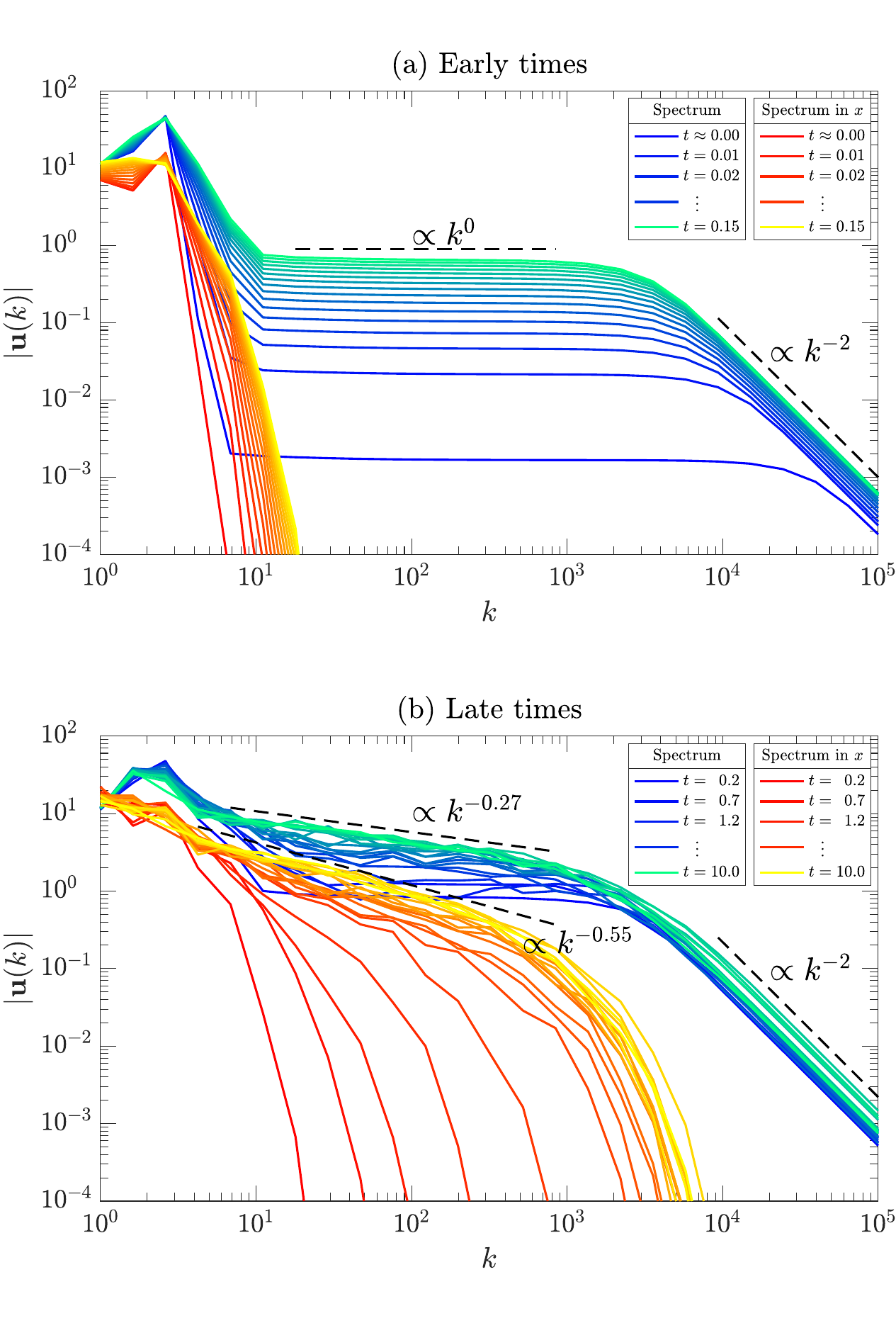}
		\caption{Velocity spectra in logarithmic scales.
			In cold colors, we plot the total spectrum~\eqref{EQ:spectrum_total}.
			In hot colors, we plot the spectrum~\eqref{EQ:spectrum_x} in $x$ direction.
			(a)~Spectra of solution for early times $t = 0, \ 0.01, \ 0.02, \ \dots,\  0.15$.
			(b)~Spectra of solution for late times $t = 0.2, \ 0.7, \ 1.2, \ \dots, \ 10.0$.}
		\label{FIG:transition}
	\end{figure*}

	Direct numerical simulations of dipole-wall collision~\cite{nguyen2011energy,nguyen2018energy} indicate that possible singularities at the boundary are expected to develop from sharp gradients in the direction tangential to the wall.
	Such sharp gradients have been related to the boundary-layer detachment.
	For this reason, we start the analysis of results by considering the simple solution spectrum
	\begin{equation}\label{EQ:spectrum_total}
		\mathcal{S}(k)= \sum_{k \leq |\mathbf{k}'| < \lambda k} |\mathbf{u}(\mathbf{k}')|
	\end{equation}
	and the spectrum in the $x$ direction
	\begin{equation}\label{EQ:spectrum_x}
		\mathcal{S}_x(k_x) = \sum_{k_y \in \mathbb{\Lambda}} |\mathbf{u}(k_x,k_y)|.
	\end{equation}
	Fig.~\ref{FIG:transition} plots the two above spectra for Run~VI, whose $Re = 10^6$.
	The behavior of early and late times are distinct.
	The initial instants in Fig.~\ref{FIG:transition}(a) depict an organized and ordered  state, followed at late times by a disorganized and chaotic state in Fig.~\ref{FIG:transition}(b).
	For this reason, we call the first regime \textit{laminar} and the second \textit{turbulent}.
	We describe their features in details now.
	
	The solutions are initialized at large scales only, but they instantly develop a $k^{-2}$ tail (which actually occurs in $k_y$ direction only, but dominates the whole spectrum) and a constant plateau $k^0$ at intermediate scales $k$.
	The explanation here of such scalings is similar to what we elaborated for one-dimensional shear flows---see the asymptotic solutions constructed in Section~\ref{SEC:pipe}.
	For large $k_y$, we have the asymptotic balance
	\begin{equation}
		\nu \partial_y^2 u \sim \nu J\delta,
	\end{equation}
	which gives us the $k_y^{-2}$ tail
	\begin{equation}
		u(k_x,k_y,t) \sim -J(k_x,t)k_y^{-2} \quad \text{for large} \ k_y.
	\end{equation}
	In intermediate scales and initial instants, we expect
	\begin{equation}
		\partial_t u \sim -\nu J \delta \quad \text{for small} \ k_y,
	\end{equation}
	whose solution provides the plateau.
	All these phenomena concern the boundary effect in $y$ direction.
	On the other hand, the spectrum in $x$ direction remains confined to large scales, slowly propagating to intermediate values of $k_x$.
	
\begin{figure*}[t]
		\centering
		\includegraphics[width=.7\textwidth]{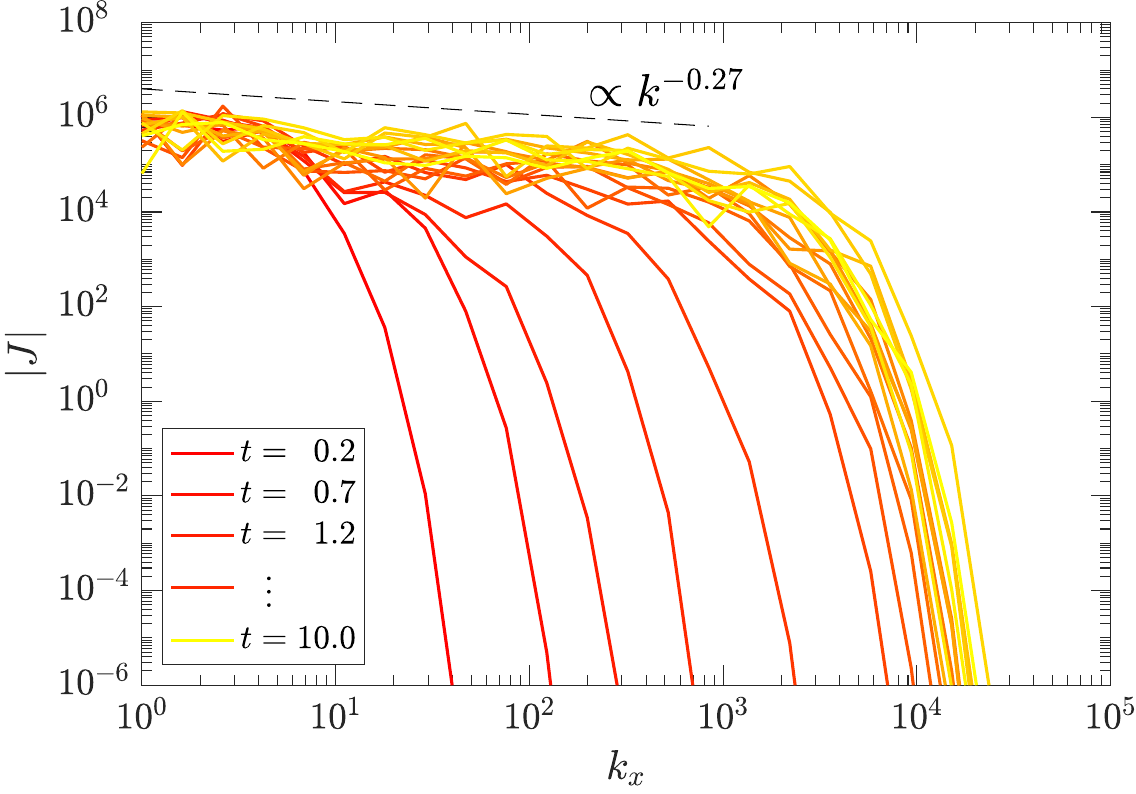}
		\caption{Jump variable spectra $J(k_x,t)$ with respect to $k_x$ in logarithmic scales.
			Time advances linearly from red to yellow as $t = 0.2, \ 0.7,\ \dots, \ 10.0$, the same instants as those in Fig.~\ref{FIG:transition}(b).}
		\label{FIG:transition_J}
	\end{figure*}

	As time advances, the plateau level increases monotonically until it reaches a certain saturated value.
	Then, other terms of the governing equation (like the nonlinear term) start to influence the solution.
	In Fig.~\ref{FIG:transition}(b) we track the abrupt transition from the laminar to turbulent states.
	The smooth and clear slopes evolve to sinuous and strong oscillations around a possible average power-law.
	The $k^{-2}$ tail is still present due to the boundary effects in $y$ direction.
	The plateau, however, changes from $k^0$ to a $k^{-0.27}$ decay.
	The $x$-spectrum $\mathcal{S}_x(k_x)$ quickly propagates towards high $k_x$ with the development of the power-law $S_x \propto k_x^{-0.55}$.
	The abrupt transition is also verified in the time evolution of the jump variable $J(k_x,t)$, depicted in Fig.~\ref{FIG:transition_J}.
	As time advances, this quantity develops the $k_x^{-0.27}$ scaling, the same observed in the total spectrum $\mathcal{S}(k)$.
	Here, the exponents are rough approximations, and they are plot in the figure just for reference.
	
	\subsection{Total dissipation at high Reynolds limit}
	
	\begin{figure*}[t]
		\centering
		\includegraphics[width=\textwidth]{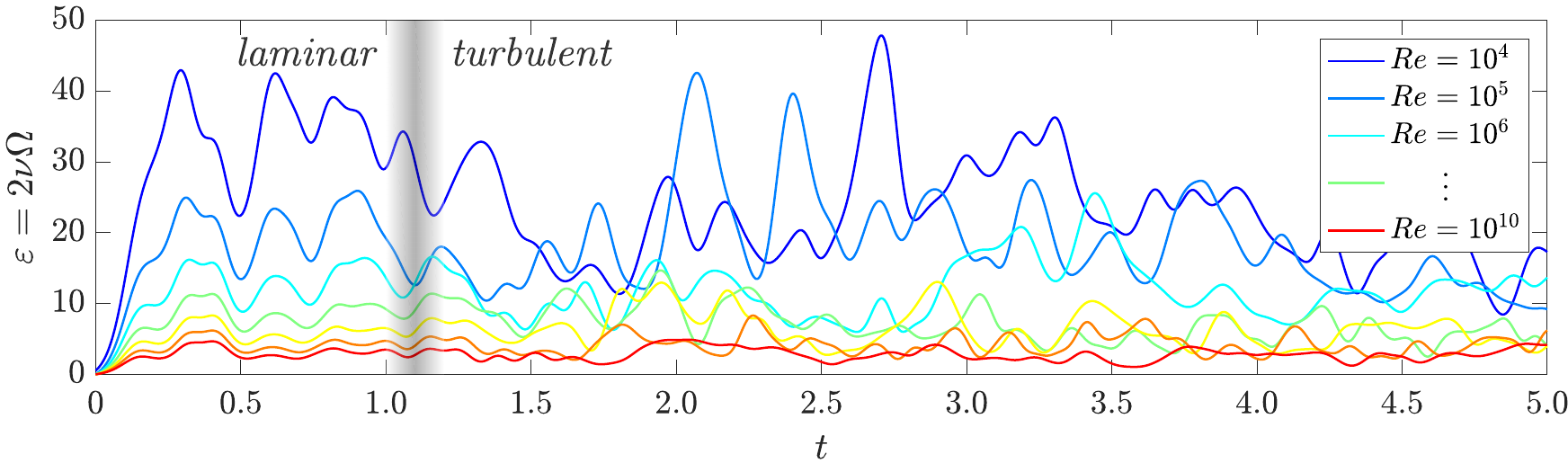}
		\caption{Time evolution of the total dissipation $\varepsilon = 2\nu\Omega$ computed from the flow enstrophy $\Omega(t)$.
			Reynolds numbers $Re = 10^4, 10^5, 10^6, \dots, 10^{10}$ increase from blue to red.
			The diffuse gray line roughly separates the laminar from the turbulent regimes.}
		\label{FIG:fig10_1}
	\end{figure*}
	
	Fig.~\ref{FIG:fig10_1} shows the time evolution of total dissipation for the different runs.
	From this picture, we can distinguish the two different regimes: at early times, when the flow is laminar, the total dissipation decreases monotonically; at later times, when it gets turbulent, humps and bumps in dissipation break monotonicity.
	The transition from an organized to a disorganized behavior is due to the appearence of chaotic oscillations in the field variables, already described in the previous section.
	The exact instant at which the transition occurs is, however, not clear from this picture.
	We shall address again this question in the next subsection.
	
	In the laminar regime, the total dissipation seems to collapse to zero in the infinite Reynolds limit.
	After the flow is turbulized, a possible convergence is not so clear, since monotonicity is broken.
	This is also the scenario in DNS~\cite{kramer2007vorticity,clercx2017dissipation}.
	Some authors conjecture the lack of convergence from Navier-Stokes to Euler in view of those bursts in dissipation after the flow turbulence is triggered~\cite{nguyen2011energy,nguyen2018energy}.
	But simulations on logarithmic lattices have the advantage of pushing the Reynolds number to incredibly large values.
	Despite the strong oscillations observed in the turbulent regime, the total dissipation appears to consistently decrease in average.
	
\begin{figure*}[t]
		\centering
		\includegraphics[width=.9\textwidth]{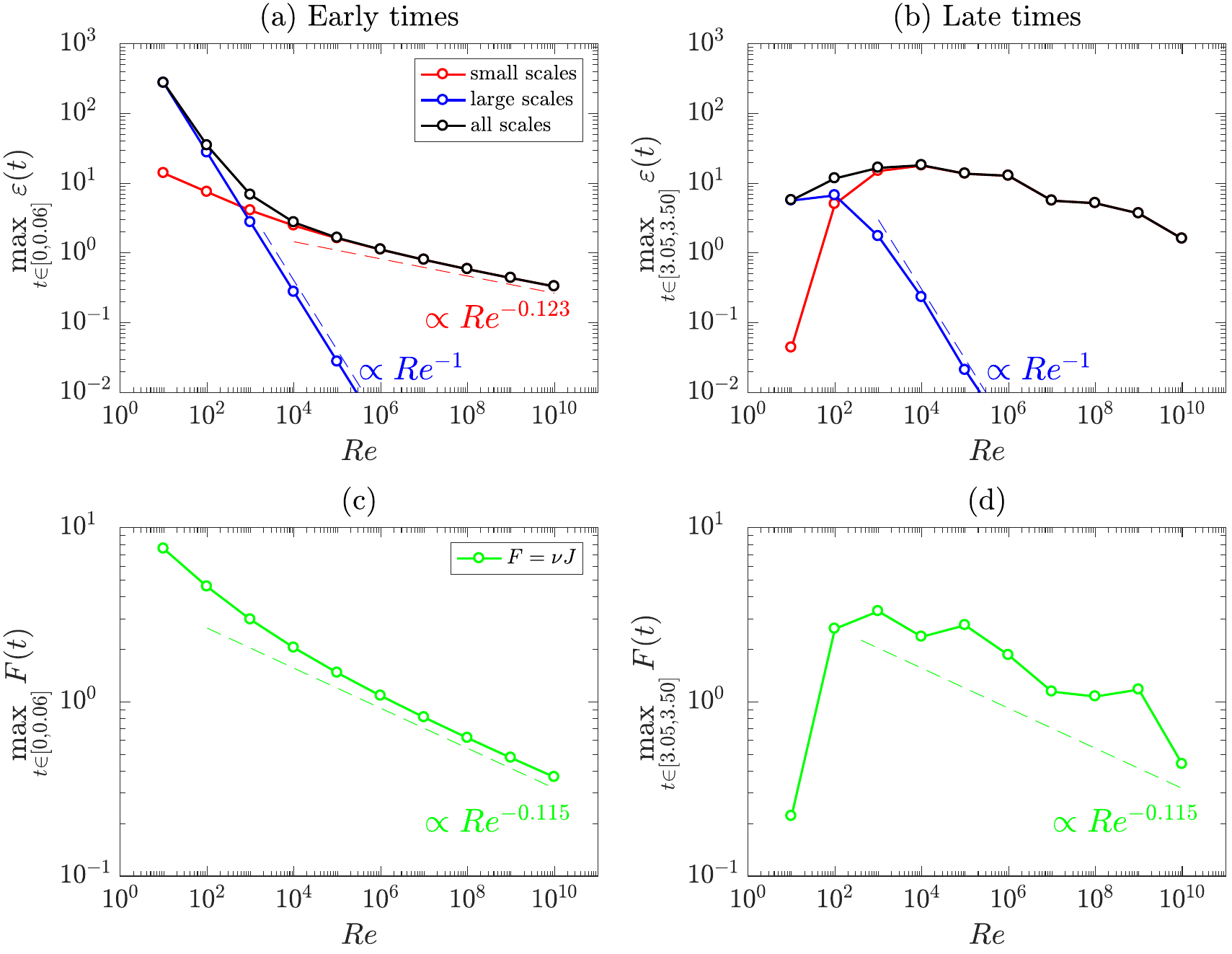}
		\caption{Scaling of some global variables with respect to Reynolds numbers.
			The first row shows maximum dissipation $\varepsilon$, and the second row maximum force intensity $F = \nu J$.
			In the left, we consider early instants of time $t \in [0,0.06]$, while in the right we study later times $t \in [3.05,3.50]$.}
		\label{FIG:fig10_4}
	\end{figure*}

	To verify how the solutions scale with Reynolds, we computed the maximum dissipation and the maximum force intensity $F = \nu J$ within two time windows: for the laminar regime, we consider the time window $t \in [0,0.06]$; for the turbulent flow, $t \in [3.05, 3.50]$.
	The results are plotted in Fig.\ref{FIG:fig10_4}.
	To explain the scaling of dissipation, we split the total dissipation into small and large scales contributions in the following way.
	Let us consider the dissipation spectrum
	\begin{equation}
		\mathcal{E}(k) = \frac{\nu}{\Delta k} \sum_{k \leq |\mathbf{k}'| < \lambda k} |k'_xk'_y|^\alpha|\pmb{\omega}(\mathbf{k}')|^{2}, \quad \text{with } \Delta k = \lambda k - k.
		\label{EQ:dissipation_spectrum}
	\end{equation}
	With this definition, the total dissipation rate $\varepsilon$ is obtained by the sum
	\begin{equation}
		\varepsilon = \sum_{1 \leq k \leq k_N} \mathcal{E}(k) \Delta k.
	\end{equation}
	The large $\mathcal{E}_l$ and small $\mathcal{E}_s$ scale contributions are expressed as
	\begin{equation}
		\mathcal{E}_l = \sum_{1 \leq k < k_{\text{cutoff}}} \mathcal{E}(k)\Delta k
		\quad \text{and} \quad
		\mathcal{E}_s = \sum_{k_{\text{cutoff} \leq k < k_{N-1}}} \mathcal{E}(k)\Delta k,
	\end{equation}
	where we set $k_{\text{cutoff}} = \lambda^5$.
	Together, they sum up the total dissipation of the flow, \textit{i.e.} $\varepsilon = \mathcal{E}_l + \mathcal{E}_s$.
	
	When the flow is still laminar, Fig.~\ref{FIG:fig10_4}(a) shows a clear convergence of dissipation towards zero.
	In the light of Kato's Theorem, the Navier-Stokes solutions are converging to the Euler's conservative solution.
	We can also study what is happening at different scales of motion.
	For smaller Reynolds numbers, the total dissipation is dominated by large scales, which decay to zero as $\propto Re^{-1}$.
	Such scaling is expected by the following reason.
	The total dissipation is given by the product $\varepsilon = 2\nu\Omega$ between the viscosity $\nu$ and the enstrophy $\Omega$.
	At large scales, turn-over time is large, so there is not much variation in the enstrophy.
	If $\Omega$ is approximately constant in this regime, the dissipation is going to decay as $\varepsilon \propto \nu = Re^{-1}$.
	We recall that this is the rate at which the Navier-Stokes solutions converge to Euler's in the absence of boundaries--- see \textit{e.g.}~\cite[Proposition~3.2]{majda2002vorticity}.
	As $Re$ increases, we have a transition around $Re = 10^3-10^4$, after which the small scales start to dominate the flow.
	In this case, the dissipation is still converging to zero, but with the slower rate $\propto Re^{-0.123}$.
	
	After the flow is turbulized, some aspects of the latter scenario change.
	We can see in Fig.~\ref{FIG:fig10_4}(b) the development of a constant plateau dissipation in intermediate values $10^3 \lessapprox Re \lessapprox 10^5$ of Reynolds number.
	The appearance of such plateau at this range of Reynolds numbers is consistent with DNS~\cite{nguyen2011energy} and it might suggest the lack of convergence to the Euler conservative solution.
	State-of-art simulations, however, cannot achieve Reynolds numbers higher than $10^5$, as it is easily reached by our logarithmic lattice model.
	By increasing $Re$, the plateau is followed by a consistent decrease in dissipation.
	This description agrees with the visual aspect of Fig.~\ref{FIG:fig10_1}, in which the total dissipation is decreasing in average.
	This suggests that dissipation must vanish in the infinite Reynolds number, but in a slow rate.
	
	Figs.~\ref{FIG:fig10_4}(c) and (d) show the corresponding scalings of the force intensity $F = \nu J$ with respect to Reynolds.
	They both decrease to zero, indicating no residual boundary shear force in the vanishing viscosity limit.
	However, the aspect of the curves differ.
	In the laminar regime, the force intensity converges to a clear power law $\propto Re^{-0.115}$, while in the turbulent regime such scaling is obeyed in average only.
	
	\subsection{Convergence in the inviscid limit}
	
	\begin{figure*}[t]
		\centering
		\includegraphics[width=\textwidth]{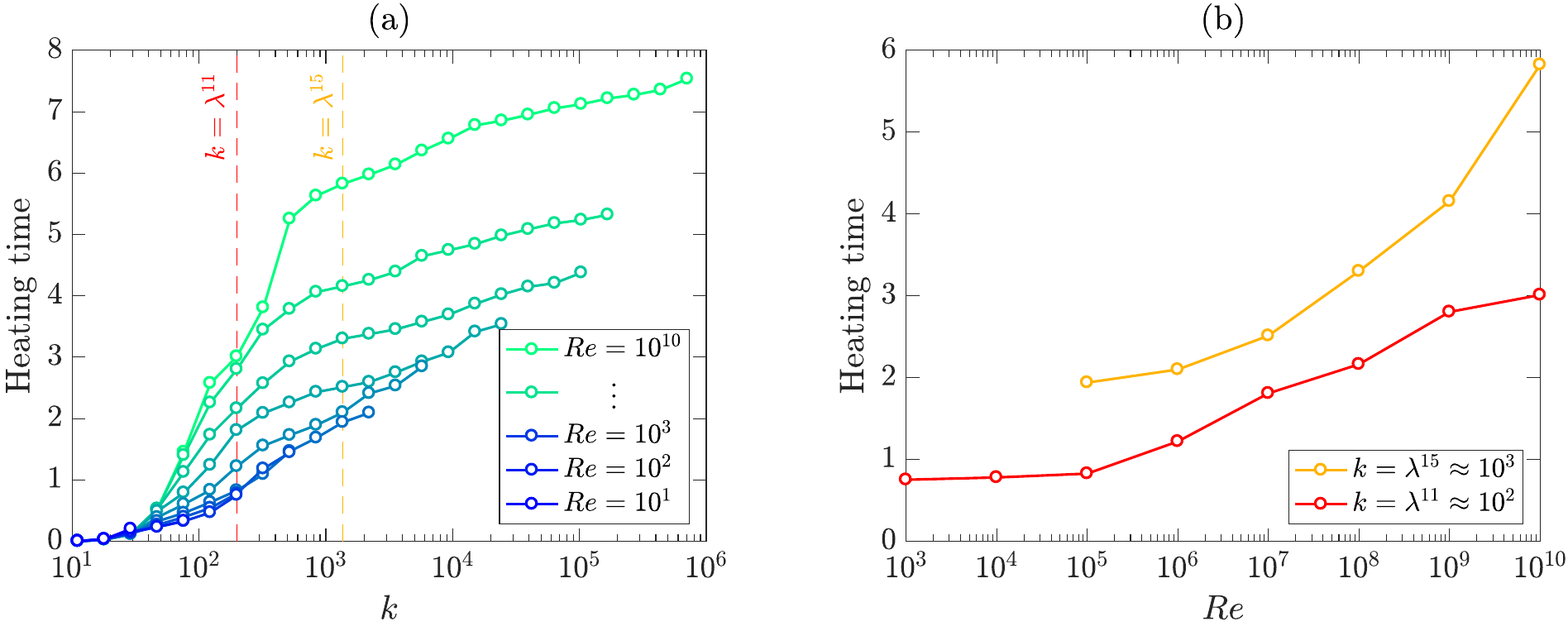}
		\caption{Heating times that each wave number $k$ requires for the spectrum $\mathcal{S}_x$ at $k$ to reach magnitude $10^{-10}$.
			(a)~Heating times of each $k$ for different Reynolds numbers, ranging from $10^1$ (in dark blue) to $10^{10}$ (in ligth green).
			(b)~Heating times against Reynolds number for the two specific wave numbers $k = \lambda^{11}$ (in red) and $k = \lambda^{15}$ (in yellow).
			These two scales are also depicted in dashed lines with their respective colors in (a) for reference.}
		\label{FIG:heating_times}
	\end{figure*}
	
	The above results suggest that dissipation vanishes as Reynolds goes to infinity.
	In view of Kato's Theorem~\cite{kato1984remarks}, this means that the Navier-Stokes solutions converge to Euler's in the vanishing viscosity limit.
	Here we address some explanations for this result.
	
	The flow is initially regular and confined to low $k_x$.
	A transition occurs, and a strong front quickly fills smaller $k_x$ scales.
	Such transition also correlates with the appearance of strong dissipation events.
	This could be interpreted as the development of sharp gradients along boundaries as the flow detaches, which is also followed by extreme events in dissipation~\cite{nguyen2011energy,nguyen2018energy,kramer2007vorticity,clercx2017dissipation}.
	Nevertheless, our computations show that the transition from laminar to turbulence and so the developement of small structures along the boundary---takes more time to take place as the Reynolds number increases.
	Indeed, Fig.~\ref{FIG:heating_times}(a) shows the heating times that each scale $k$ needs for the spectrum $\mathcal{S}_x(k)$ at $k$ to reach magnitude $10^{-10}$.
	As we can see, the heating times increase with the Reynolds number, a fact which is confirmed in Fig.~\ref{FIG:heating_times}(b) for two fixed scales $k = \lambda^{11}$ and $k = \lambda^{15}$.
	This explains why, in a fixed interval of time, the dissipation looks like is converging to a constant value, but start vanishing again with increasing Reynolds: the transition instants drift away from the time window for larger Reynolds numbers.
	
	In the case of dipole-wall collision, we don't expect the instant of transition to advance as the Reynolds number increases.
	Indeed, turbulence is triggered by the collision of the vortex against the wall, and the dipole travel time is supposedly governed by the Reynolds-independent bulk flow~\cite{orlandi1990vortex}.
	We remark, however, that the results on logarithmic lattices do not disagree with those from DNS~\cite{nguyen2011energy,nguyen2018energy}, if we compare total dissipation in their common Reynolds number range.
	For instance, if we look at the Reynolds interval $10^3 \leq Re \leq 10^5$, the dissipation in the logarithmic lattice model from Fig.~\ref{FIG:fig10_4}(b) is similar to DNS results from~\cite{nguyen2011energy}.
	
	As we mentioned in the description of the numerical setup, we performed some of the simulations with twenty less and twenty more nodes, in order to check the influence of lattice truncation on the results.
	The same qualitative picture was observed for all runs.
	We believe, however, that more rigorous numerical study should be undertaken.
	
	\section{Conclusions}\label{SEC:CONC}
	
	We showed that logarithmic lattices are able to model systems with solid boundaries.
	We explored several examples of shear flows to create intuition on the unusual description of flows with walls in Fourier space.
	The vanishing viscosity problem was approached for two-dimensional flows in the presence of a flat solid boundary.
	We tracked total dissipation, in the light of Kato's Theorem.
	As time evolves, a laminar organized regime is followed by a chaotic turbulent state.
	This resembles the development of small-scale structures in the dipole-wall collision.
	Nevertheless, the Navier-Stokes solutions seem to converge to Euler's in the infinite Reynolds limit.
	We explained that the propagation of energy from large to small scales in $k_x$ takes more and more time with increasing Reynolds, which justifies the vanishing dissipation on compact time intervals.
	We argued that this fact does not contradict the reported DNS from literature, since the results are in agreement in the resolution accessible to the state-of-the-art DNS.
	Further investigation, however, is needed to obtain a better understanding of this challenging problem.
	Still, the logarithmic lattice model looks like a promising tool for this and other related problems, since it reaches the large Reynolds number $Re = 10^{10}$ with moderate computational effort compared to the expensive DNS~\cite{nguyen2011energy,clercx2017dissipation,nguyen2018energy} limited to $Re = 10^5$.
	
	We expect that the techniques presented here will be useful for the study of many relevant problems surrounding the effects of boundaries.
	This includes the boundary layer instabilities~\cite{grenier2000nonlinear}, the finite-time blowup in the Prandtl equations~\cite{weinan1997blowup,kukavica2017van} and further studies about the inviscid limit of the Navier-Stokes equations~\cite{weinan2000boundary}.
	Another problem that fits our framework is the developement of finite-time singularities in the Euler equations with boundaries~\cite{kiselev2018small}, which, despite the recent successful mathematical advances~\cite{elgindi2019finite,chen2021finite} and the strong numerical evidences~\cite{luo2013potentially}, is still under active investigation~\cite{kolluru2022insights}.
	
	The approach presented in this paper is restricted to a steady flat solid boundary.
	While this is a good setup for the study of local singularities, more general problems may involve complex geometries, unsteady walls or even multiple boundaries.
	We leave these interesting but not straightforward generalizations to future works.

	\section*{Acknowledgments}
	
	The authors express their gratitude to B\'{e}reng\`{e}re Dubrulle and Simon Thalabard for their continuous support and debate on logarithmic lattices.
	They are also grateful to Marie Farge for the insights about her DNS results on the vanishing viscosity limit.
	C.C. was supported by the French National Research Agency (ANR project TILT; ANR-20-CE30-0035).
	A.M. was supported by CNPq grant 308721/2021-7, FAPERJ grant E-26/201.054/2022, and FAPERJ Pensa Rio grant E-26/210.874/2014.
	C.C.~acknowledges the finantial support from IMPA for his visit, where the writing of this manuscript has been completed.
	C.C.~dedicates this work to Our Lady of Mount Carmel.
	
	\section*{Appendix A: Discontinuous Navier-Stokes equations}\label{app:A}
	
	In this appendix, we provide some details and references about the Navier-Stokes equations for flows with discontinuities.
	
	Discontinuous formulations of the Fluid Dynamics equations were first derived in~\cite{Sirovich1967Ini} for compressible ideal flows and subsequently applied to gas dynamics~\cite{Sirovich1968Ste}, compressible magnetohydrodynamics~\cite{Salathe1967Bou}, and in the modelling of no-slip boundary for incompressible viscous flow~\cite{Goldstein1993Mod}.
	It also inspired many immersed boundary techniques that revealed useful for the numerical simulation of complex boundary geometries~\cite{peskin1972flow,peskin2002immersed}.
	
	Let us allow the fields to have jump discontinuities on a surface $\mathcal{S}_t$ immersed on the flow.
	Assuming the flow to be incompressible and that there is no penetration accross $\mathcal{S}_t$, the governing equations are
	\begin{equation}
		\partial_t \mathbf{u} + \mathbf{u} \cdot \nabla \mathbf{u} = -\nabla p + \nu \Delta \mathbf{u} + [\pmb{\sigma} \cdot \mathbf{n}]\delta_{\mathcal{S}_t}, \quad \nabla \cdot \mathbf{u} = 0,
		\label{EQ:NS_disc}
	\end{equation}
	where $\pmb{\sigma}$ is the stress tensor, $\mathbf{n}$ is the unit normal vector to $\mathcal{S}_t$ and the brackets $[\cdot]$ indicate the jump accross $\mathcal{S}_t$.
	The convention for the jumps is the following: if $\mathcal{S}_t$ splits the domain into two subdomains---say a positive and a negative one---then $\mathbf{n}$ points from the positive towards the negative, and the jump is computed as $[f] = f^+-f^-$, where $f^+$ is the value at the positive side, and $f^-$ at the negative side.
	Finally, we have also introduced the Dirac delta $\delta_{\mathcal{S}_t}$ locating $\mathcal{S}_t$ given by
	\begin{equation}
		\delta_{\mathcal{S}_t}(\mathbf{x}) = \int_{\mathcal{S}_t} \prod_{i = 1}^{d} \delta(x_i - y_i)dS(\mathbf{y}),
		\label{EQ:dirac}
	\end{equation}
	where $\delta$ is the usual Dirac delta, and $d$ is the spatial dimension.
	
	In the framework under consideration in this paper, we can compute the term $[\pmb{\sigma} \cdot \mathbf{n}] \delta_{\mathcal{S}_t}$ explicitly.
	Our boundary is the steady plane $y=0$, so we have simply $\delta_{\mathcal{S}_t} = \delta(y)$.
	It remains to establish the jump $[\pmb{\sigma}\cdot \mathbf{n}]$ of tractions accross the boundary.
	First, we set the upper-half space $y > 0$ as being the positive domain split by the discontinuity surface $y = 0$, while the lower-half space $y < 0$ is its negative counterpart.
	Following the convention established above, the unit normal vector at $y = 0$ pointing from the positive towards the negative domains is $\mathbf{n} = (0,-1,0)^T$.
	Next, the stress tensor on $y = 0$ is given by
	\begin{equation}
		\pmb{\sigma} =
		\begin{pmatrix}
			-p & \nu\partial_y u & 0\\
			\nu\partial_y u & -p + 2\nu\partial_y v & \nu \partial_y w\\
			0 & \nu \partial_y w & -p
		\end{pmatrix}
		\quad \text{on} \ y = 0,
	\end{equation}
	where we have used the no-slip boundary condition $\mathbf{u} = \mathbf{0}$ on the $x$-$z$ plane to neglect all terms involving spatial derivatives in $x$ and $z$ directions, that is $\partial_x u = \partial_x v = \partial_x w = \partial_z u = \partial_z v = \partial_z w = 0$ on $y = 0$.
	Hence, 
	\begin{equation}
		\pmb{\sigma} \cdot \mathbf{n} =
		\begin{pmatrix}
			-\nu\partial_y u\\
			p - 2\nu\partial_y v\\
			-\nu \partial_y w
		\end{pmatrix}
		\quad
		\text{on} \ y = 0.
	\end{equation}
	Finally, because of the symmetries~\eqref{EQ:flow_symmetry} on $p$ and $v$, we have $[p] = [\partial_y v] = 0$, so we obtain
	\begin{equation}\label{EQ:jumps_partial}
		[\pmb{\sigma} \cdot \mathbf{n}] =
		-\nu
		\begin{pmatrix}
			[\partial_y u]\\
			0\\
			[\partial_y w]
		\end{pmatrix}.
	\end{equation}
	This leads to system~\eqref{EQ:flat_plate_model}-\eqref{EQ:J}.
	
	\section*{Appendix B: Boundary conditions and nonuniqueness of immersed boundary flows}\label{app:B}
	
	The purpose of this appendix is to show that we cannot suppress the boundary condition in the immersed boundary models, otherwise the problem becomes ill-posed.
	More precisely, we show that if we do not impose the no-slip condition, solutions are nonunique.
	For sake of simplicity, we do it for one-dimensional shear flows.
	
	Let us consider the initial value problem
	\begin{equation}
		\begin{cases}
			\partial_t u = \nu \partial_y^2 u - \nu J(t) \delta(y) \quad &\text{in} \ y \in \mathbb{R}, \ t>0\\
			u(y,t) = u(-y,t) \quad &\text{in} \ y \in \mathbb{R}, \ t > 0\\
			u(y,0) = g(y) \quad &\text{for} \ y \in \mathbb{R}.
			\label{EQ:simpler_system_disc_}
		\end{cases}
	\end{equation}
	with the jump $J(t)$ given by
	\begin{equation}
		J(t) = [\partial_yu].
	\end{equation}
	This is the immersed boundary formulation of the one-dimensional unforced shear flow~\eqref{EQ:simpler_system_control}, but here the boundary condition $u = 0$ at $y=0$ is not imposed.
	Using the classical theory of heat equation, we shall construct two distinct solutions for problem~\eqref{EQ:simpler_system_disc_}.
	
	Let us define
	\begin{equation}\label{EQ:nonunique1}
		u(y,t) = \int_{0}^{\infty} \left( K(y,x,t) - K(y,-x,t) \right)g(x)dx, \quad \text{for} \ y>0
	\end{equation}
	and $u(-y,t) = u(y,t)$ for all $y \in \mathbb{R}$, where
	\begin{equation}
		K(y,x,t) = \frac{e^{-|y-x|^2/4\nu t}}{(4\pi \nu t)^{1/2}}
	\end{equation}
	is the heat kernel~\cite{John1978partial}.
	We claim that such function solves problem~\eqref{EQ:simpler_system_disc_}.
	Indeed, \eqref{EQ:nonunique1} is the solution of the Dirichlet problem
	\begin{equation}\label{EQ:heat_dirichlet}
		\begin{cases}
			\partial_t u = \nu \partial_y^2 u \quad &\text{in} \ y>0, \ t>0\\
			u(y,0) = g(y) \quad &\text{for} \ y >0\\
			u(0,t) = 0 \quad &\text{for} \ t>0 \ \text{(Dirichlet B.C.)}.
		\end{cases}
	\end{equation}
	Therefore, the constructed $u(y,t)$ is the physical solution of problem~\eqref{EQ:simpler_system_disc_}, \textit{i.e.} the solution which satisfies the no-slip condition $u=0$ at $y=0$.
	
	\begin{figure*}[t]
		\centering
		\includegraphics[width=\textwidth]{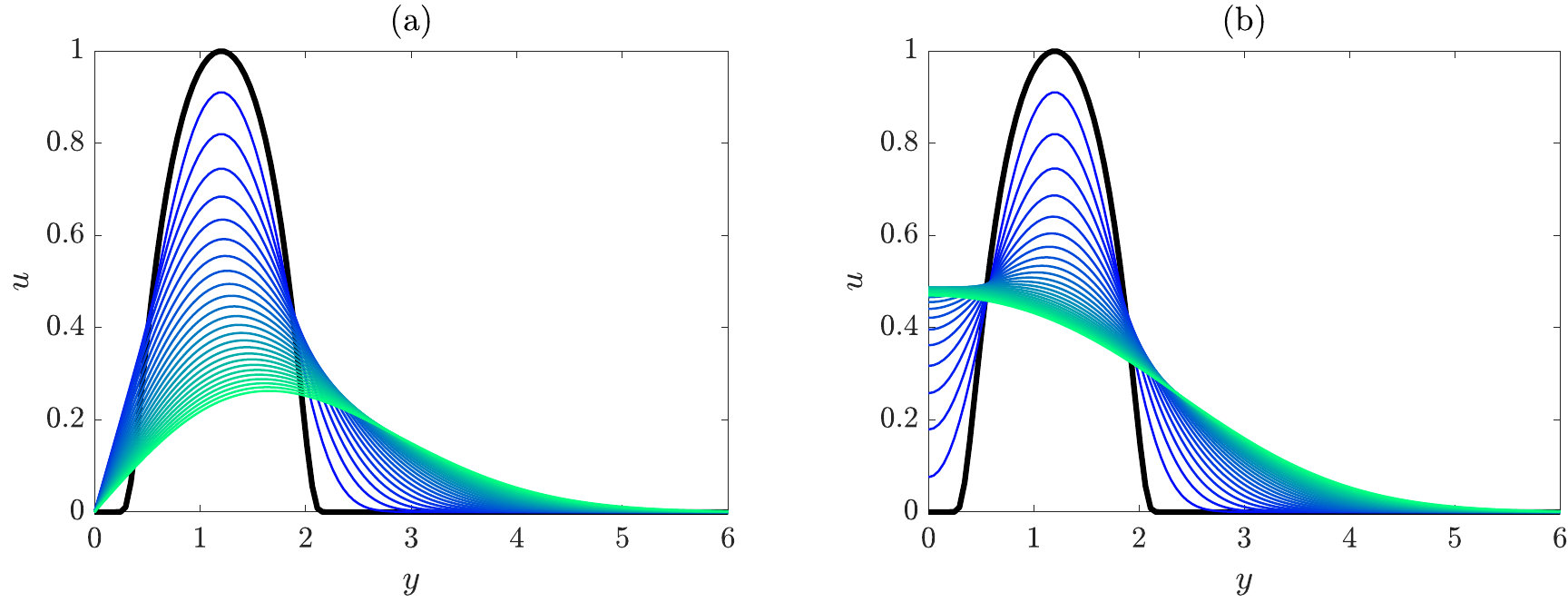}
		\caption{Two distinct solutions for the immersed boundary model~\eqref{EQ:simpler_system_disc_} from the same intial condition (in black). Initial data satisfy no-slip $u=0$ at $y=0$. Colors change from blue to green as time advances. (a) solution with no-slip at the boundary, obtained from imposing Dirichlet boundary condition $u = 0$ at $y= 0$; (b) solution with slip at the boundary, satisfying the Neumann boundary condition $\partial_y u = 0$ at $y = 0$.}
		\label{FIG:two_solutions}
	\end{figure*}
	
	Next, define
	\begin{equation}\label{EQ:nonunique2}
		u(y,t) = \int_{0}^{\infty} \left( K(y,x,t) + K(y,-x,t) \right)g(x)dx, \quad \text{for} \ y>0
	\end{equation}
	and $u(-y,t) = u(y,t)$ for all $y \in \mathbb{R}$.
	We claim that this is also a solution of problem~\eqref{EQ:simpler_system_disc_}.
	Indeed, \eqref{EQ:nonunique2} is the solution of the Neumann problem
	\begin{equation}\label{EQ:heat_neumann}
		\begin{cases}
			\partial_t u = \nu \partial_y^2 u \quad &\text{in} \ y>0, \ t>0\\
			u(y,0) = g(y) \quad &\text{for} \ y >0\\
			\partial_yu(0,t) = 0 \quad &\text{for} \ t>0 \ \text{(Neumann B.C.)}.
		\end{cases}
	\end{equation}
	Therefore, the constructed $u(y,t)$ by the reflection trivially satisfies problem~\eqref{EQ:simpler_system_disc_}, since the jump variable is identically zero $J \equiv 0$.
	So we have constructed two distinct solutions to for problem~\eqref{EQ:simpler_system_disc_}.
	
	In Fig.~\ref{FIG:two_solutions} we show the two distinct solutions constructed above for the initial condition
	\begin{equation}
		g(y) = C \exp\left( -\frac{1}{1-|y-y_0|^2} \right),
	\end{equation}
	with $y_0 = 1.2$ and $C$ such that $\max_{y} g(y) = 1$.
	Such function has compact support $\text{supp} \ g = [0.2,2.2]$; particularly, $g(0) = 0$.
	With this example, we also show that it is not sufficient to have no-slip condition at the initial instant for the solution to satisy no-slip at all subsequent instants.
	In Fig.~\ref{FIG:two_solutions}(a) we see the physical expected solution, which satisfies no-slip condition at all times.
	Differently, Fig.~\ref{FIG:two_solutions}(b) shows a solution which initially satisfies $u=0$ at $y=0$, but develops non zero slip as time advances.
	
	\bibliographystyle{plain}
	\bibliography{refs}

\begin{thebibliography}{10}

\bibitem{bardos2013mathematics}
C.~W. Bardos and E.~S. Titi.
\newblock {Mathematics and turbulence: where do we stand?}
\newblock {\em J. Turbul.}, 14(3):42--76, 2013.

\bibitem{barral2023asymptotic}
A.~Barral and B.~Dubrulle.
\newblock {Asymptotic ultimate regime of homogeneous Rayleigh--B{\'e}nard
  convection on logarithmic lattices}.
\newblock {\em J. Fluid Mech.}, 962:A2, 2023.

\bibitem{biferale2003shell}
L.~Biferale.
\newblock Shell models of energy cascade in turbulence.
\newblock {\em Ann. Rev. Fluid Mech.}, 35:441--468, 2003.

\bibitem{campolina2019fluid}
C.~S. Campolina.
\newblock {Fluid Dynamics on Logarithmic Lattices and Singularities of Euler
  Flow}.
\newblock {Master's Thesis}, Instituto de Matem\'{a}tica Pura e Aplicada, 2019.

\bibitem{campolina2020loglatt}
C.~S. Campolina.
\newblock Loglatt: A computational library for the calculus and flows on
  logarithmic lattices.
\newblock {\em arXiv preprint arXiv:2006.00047}, 2020.

\bibitem{campolina2020loglattmatlab}
C.~S. Campolina.
\newblock {LogLatt: A computational library for the calculus on logarithmic
  lattices}, 2020.
\newblock Freely available for noncommercial use from MATLAB Central File
  Exchange
  (\url{https://www.mathworks.com/matlabcentral/fileexchange/76295-loglatt}).

\bibitem{campolina2022fluid}
C.~S. Campolina.
\newblock {Fluid Flows and Boundaries on Logarithmic Lattices}.
\newblock {PhD Thesis}, Instituto de Matem\'{a}tica Pura e Aplicada, 2022.

\bibitem{campolina2018chaotic}
C.~S. Campolina and A.~A. Mailybaev.
\newblock {Chaotic blowup in the 3D incompressible Euler equations on a
  logarithmic lattice}.
\newblock {\em Phys. Rev. Lett.}, 121:064501, 2018.

\bibitem{campolina2021fluid}
C.~S. Campolina and A.~A. Mailybaev.
\newblock Fluid dynamics on logarithmic lattices.
\newblock {\em Nonlinearity}, 34(7):4684, 2021.

\bibitem{cassel2000comparison}
K.~W. Cassel.
\newblock {A comparison of Navier-Stokes solutions with the theoretical
  description of unsteady separation}.
\newblock {\em Philos. Trans. Roy. Soc. A}, 358(1777):3207--3227, 2000.

\bibitem{chen2021finite}
J.~Chen and T.~Y. Hou.
\newblock {Finite time blowup of 2D Boussinesq and 3D Euler equations with
  {$C^{1,\alpha}$} velocity and boundary}.
\newblock {\em Comm. Math. Phys.}, 383(3):1559--1667, 2021.

\bibitem{clercx2017dissipation}
H.~J.~H. Clercx and G.~J.~F. Van~Heijst.
\newblock Dissipation of coherent structures in confined two-dimensional
  turbulence.
\newblock {\em Physics of Fluids}, 29(11):111103, 2017.

\bibitem{costa2023reversible}
G.~Costa, A.~Barral, and B.~Dubrulle.
\newblock {Reversible Navier-Stokes equation on logarithmic lattices}.
\newblock {\em Phys. Rev. E}, 107(6):065106, 2023.

\bibitem{weinan2000boundary}
W.~E.
\newblock {Boundary layer theory and the zero-viscosity limit of the
  Navier-Stokes equation}.
\newblock {\em Acta Math. Sin.}, 16:207--218, 2000.

\bibitem{weinan1997blowup}
W.~E and B.~Engquist.
\newblock {Blowup of solutions of the unsteady Prandtl’s equation}.
\newblock {\em Comm. Pure Appl. Math}, 50(12):1287--1293, 1997.

\bibitem{elgindi2019finite}
T.~M. Elgindi and I.-J. Jeong.
\newblock {Finite-time singularity formation for strong solutions to the
  axi-symmetric 3D Euler equations}.
\newblock {\em Ann. PDE}, 5(2):16, 2019.

\bibitem{friedlander1998introduction}
F.~G. Friedlander.
\newblock {\em {Introduction to the Theory of Distributions}}.
\newblock Cambridge University Press, 1998.

\bibitem{gerard2010ill}
D.~G{\'e}rard-Varet and E.~Dormy.
\newblock {On the ill-posedness of the Prandtl equation}.
\newblock {\em J. Amer. Math. Soc.}, 23(2):591--609, 2010.

\bibitem{Goldstein1993Mod}
D.~Goldstein, R.~Handler, and L.~Sirovich.
\newblock Modeling a no-slip flow boundary with an external force field.
\newblock {\em Journal of Computational Physics}, 105(2):354--366, 1993.

\bibitem{grenier2000nonlinear}
E.~Grenier.
\newblock {On the nonlinear instability of Euler and Prandtl equations}.
\newblock {\em Comm. Pure Appl. Math.}, 53(9):1067--1091, 2000.

\bibitem{John1978partial}
F.~John.
\newblock {\em {Partial Differential Equations}}, volume~1 of {\em {Applied
  Mathematical Sciences}}.
\newblock Springer, 3rd edition, 1978.

\bibitem{kato1972nonstationary}
T.~Kato.
\newblock {Nonstationary flows of viscous and ideal fluids in {$R^3$}}.
\newblock {\em J. Funct. Anal.}, 9(3):296--305, 1972.

\bibitem{kato1984remarks}
T.~Kato.
\newblock {Remarks on zero viscosity limit for nonstationary Navier-Stokes
  flows with boundary}.
\newblock In {\em Seminar on nonlinear partial differential equations}, pages
  85--98. Springer, 1984.

\bibitem{kiselev2018small}
A.~Kiselev.
\newblock {Small scales and singularity formation in Fluid Dynamics}.
\newblock In {\em Proc. Int. Cong. of Math. 2018}, volume~2, pages 2329--2356,
  Rio de Janeiro, 2018. World Scientific.

\bibitem{kolluru2022insights}
S.~S.~V. Kolluru, P.~Sharma, and R.~Pandit.
\newblock {Insights from a pseudospectral study of a potentially singular
  solution of the three-dimensional axisymmetric incompressible Euler
  equation}.
\newblock {\em Phys. Rev. E}, 105(6):065107, 2022.

\bibitem{kramer2007vorticity}
W.~Kramer, H.~J.~H. Clercx, and G.~J.~F. Van~Heijst.
\newblock Vorticity dynamics of a dipole colliding with a no-slip wall.
\newblock {\em Phys. Fluids}, 19(12):126603, 2007.

\bibitem{kukavica2017van}
I.~Kukavica, V.~Vicol, and F.~Wang.
\newblock {The van Dommelen and Shen singularity in the Prandtl equations}.
\newblock {\em Adv. Math.}, 307:288--311, 2017.

\bibitem{landau1987fluid}
L.~D. Landau and E.~M. Lifshitz.
\newblock {\em {Fluid Mechanics}}, volume~6.
\newblock Pergamon, 1987.

\bibitem{leveque2007finite}
R.~J. LeVeque.
\newblock {\em {Finite difference methods for ordinary and partial differential
  equations: steady-state and time-dependent problems}}.
\newblock SIAM, 2007.

\bibitem{lopes2018fluids}
H.~J.~N. Lopes and M.~C.~Lopes Filho.
\newblock Fluids, walls and vanishing viscosity.
\newblock In {\em Proc. Int. Cong. Math. 2018}, volume~3, pages 2537--2558, Rio
  de Janeiro, 2018. World Scientific.

\bibitem{luo2013potentially}
G.~Luo and T.~Y. Hou.
\newblock {Potentially singular solutions of the 3D incompressible Euler
  equations}.
\newblock {\em PNAS}, 111:12968--1297349, 2014.

\bibitem{majda2002vorticity}
A.~J. Majda and A.~L. Bertozzi.
\newblock {\em {Vorticity and Incompressible Flow}}.
\newblock Cambridge University Press, 2002.

\bibitem{nguyen2018energy}
N.~Nguyen~van yen, M.~Waidmann, R.~Klein, M.~Farge, and K.~Schneider.
\newblock Energy dissipation caused by boundary layer instability at vanishing
  viscosity.
\newblock {\em J. Fluid Mech.}, 849:676--717, 2018.

\bibitem{nguyen2011energy}
R.~Nguyen~van yen, M.~Farge, and K.~Schneider.
\newblock Energy dissipating structures produced by walls in two-dimensional
  flows at vanishing viscosity.
\newblock {\em Phys. Rev. Lett.}, 106:184502, 2011.

\bibitem{orlandi1990vortex}
P.~Orlandi.
\newblock Vortex dipole rebound from a wall.
\newblock {\em Physics of Fluids A: Fluid Dynamics}, 2(8):1429--1436, 1990.

\bibitem{peskin1972flow}
C.~S. Peskin.
\newblock Flow patterns around heart valves: a numerical method.
\newblock {\em Journal of computational physics}, 10(2):252--271, 1972.

\bibitem{peskin2002immersed}
C.~S. Peskin.
\newblock The immersed boundary method.
\newblock {\em Acta numerica}, 11:479--517, 2002.

\bibitem{pikeroen2023tracking}
Q.~Pikeroen, A.~Barral, G.~Costa, C.~Campolina, A.~Mailybaev, and B.~Dubrule.
\newblock Tracking complex singularities of fluids on log-lattices.
\newblock {\em Nonlinearity (submitted)}, 2023.

\bibitem{pikeroen2023log}
Q.~Pikeroen, A.~Barral, G.~Costa, and B.~Dubrulle.
\newblock Log-lattices for atmospheric flows.
\newblock {\em Atmosphere}, 14(11):1690, 2023.

\bibitem{Salathe1967Bou}
E.~P. Salath\'{e} and L.~Sirovich.
\newblock Boundary-value problems in compressible magnetohydrodynamics.
\newblock {\em The Physics of Fluids}, 10(7):1477--1491, 1967.

\bibitem{schlichting1961boundary}
H.~Schlichting and J.~Kestin.
\newblock {\em {Boundary Layer Theory}}, volume 121.
\newblock Springer, 1961.

\bibitem{shampine1997matlab}
L.~F. Shampine and M.~W. Reichelt.
\newblock {The Matlab ODE suite}.
\newblock {\em SIAM J. Sci. Comput.}, 18(1):1--22, 1997.

\bibitem{Sirovich1967Ini}
L.~Sirovich.
\newblock Initial and boundary value problems in dissipative gas dynamics.
\newblock {\em The Physics of Fluids}, 10(1):24--34, 1967.

\bibitem{Sirovich1968Ste}
L.~Sirovich.
\newblock Steady gasdynamic flows.
\newblock {\em The Physics of Fluids}, 11(7):1424--1439, 1968.

\bibitem{trefethen2000spectral}
L.~N. Trefethen.
\newblock {\em {Spectral methods in MATLAB}}.
\newblock SIAM, 2000.

\end{thebibliography}

\end{document}